\documentclass[fleqn,usenatbib]{mnras}
\usepackage{newtxtext}
\usepackage[varg]{newtxmath}    
\usepackage[T1]{fontenc}    
\usepackage{ae,aecompl}    
\usepackage{graphicx}
\usepackage{amsmath}	
\usepackage{amssymb}	
\usepackage{color}
\usepackage{physics}	
\usepackage{multirow}
\usepackage{journals}
\usepackage{bm}	
\usepackage{url}
\usepackage{footnote}
\makesavenoteenv{tabular}
\makesavenoteenv{table}
\definecolor{darkgreen}{rgb}{0.0,0.5,0.0}
\definecolor{darkred}{rgb}{0.5,0.0,0.0}
\definecolor{brown}{rgb}{0.65,.16,0.16}
\definecolor{grey}{rgb}{0.4,0.5,0.6}

\newcommand{\deltam}{$\Delta \mathcal{M}_{12}$}

\newcommand{\Mabs}{$\mathcal{M}_{r}^{\rm Petro}$}
\newcommand{\mr}{$m_{r}^{\rm Petro}$}
\newcommand{\Mvir}{$\log(M_{\rm vir}/{\rm M_{\odot}})$}
\newcommand{\MvirNoUnit}{$\log M_{\rm vir}$}

\newcommand{\Rmax}{$R_{\rm max}$}
\newcommand{\Rvir}{$r_{\rm vir}$}
\newcommand{\halfRvir}{$0.5\,r_{\rm vir}$}

\title[CLFs and magnitude-gap statistics]
{A finer view of the conditional galaxy luminosity function and magnitude-gap statistics}
\author[M. Trevisan \& G. A. Mamon]{M. Trevisan$^{1}$\thanks{E-mail: trevisan@iap.fr}
and
G. A. Mamon$^{1}$ \\
 $^{1}$Institut d'Astrophysique de Paris (UMR 7095: CNRS \& UPMC, Sorbonne Universit\'es), 98 bis Bd Arago, 75014 Paris, France 
}
 
\begin{document}

\date{Accepted ---. Received ---; in original form ---}

\pagerange{\pageref{firstpage}--\pageref{lastpage}} \pubyear{2016}

\maketitle

\label{firstpage}

\begin{abstract}
The gap between first and second ranked galaxy magnitudes in groups is often considered a tracer of their merger histories, which in turn may affect galaxy properties, and also serves to test galaxy luminosity functions (LFs).
We remeasure the conditional luminosity function (CLF) of the Main Galaxy Sample of the SDSS  in an appropriately cleaned subsample of groups from the Yang catalog. We find that, at low group masses, our best-fit CLF have steeper satellite high ends, yet higher ratios of characteristic satellite to central luminosities in comparison  with the CLF of Yang et al. (2008).
The observed fractions of groups with large and small magnitude gaps 
as well as the Tremaine \& Richstone (1977) statistics,
are not compatible with either a single Schechter LF or with a  Schechter-like satellite plus lognormal central LF.
These gap statistics, which naturally depend on the size of the subsamples, and also on the maximum projected radius, $R_{\rm max}$, for defining the 2nd brightest galaxy, can only be reproduced with two-component CLFs  if we allow small gap groups to preferentially have two central galaxies, as expected when groups merge. 
Finally, we find that the trend of higher gap for higher group velocity dispersion, $\sigma_{\rm v}$, at given richness, discovered by Hearin et al. (2013), is strongly reduced when we consider $\sigma_{\rm v}$ in bins of richness, and virtually disappears when we use group mass instead of $\sigma_{\rm v}$.
This limits the applicability of gaps in refining cosmographic studies based on cluster counts.
\end{abstract}


\section{Introduction}
\label{Sec_intro}

One of the most fundamental properties of galaxy populations are their distributions of stellar masses and luminosities. Indeed, as galaxies form in dark matter haloes, their stellar mass function (SMF) and luminosity function (LF) combine information on the halo mass function. Moreover, galaxies evolve in many ways, through both internal  processes as well as environmental processes, and these leave imprints in the SMF and LF. For example, the lower mass groups of galaxies (excluding the higher mass clusters) are the primary sites of galaxy mergers (e.g., \citealt{Mamon:1992}). 

In a series of important articles, Yang and co-authors have modified our view of the galaxy SMF and LF.
\cite{Yang.etal:2007} first designed a new type of group finder that extracts groups from galaxy catalogues in redshift space. In their group finder, the most luminous galaxies or the most massive in stars (hereafter called \emph{central} galaxies) have a special role in the group, as in most semi-analytical models of galaxy formation \citep{Kauffmann.etal:1999,Croton.etal:2006,deLucia.Blaizot:2007,Guo.etal:2011,Henriques.etal:2015}.
Testing group finders on mock catalogues,
\cite{Duarte.Mamon:2015} found that both their own group finder MAGGIE and the Yang et al. group finder  are much more accurate than the popular \emph{Friends-of-Friends} (FoF) method \citep{Huchra.Geller:1982,Berlind.etal:2006,Robotham.etal:2011}. Indeed, the groups extracted with  MAGGIE and the \cite{Yang.etal:2007} group algorithm are much less prone to be secondary fragments of more massive groups and lead to much more accurate and less biased group masses than one infers for FoF groups using the virial theorem.

In subsequent work, \citeauthor*{Yang.etal:2008} (\citeyear{Yang.etal:2008}, hereafter Y08) have shown that the galaxy LF strongly depends on group mass, in particular the characteristic luminosity at the knee of the LF increases with group mass. Y08 thus measured the \emph{conditional luminosity function} (CLF). Moreover, Y08  established that in sufficiently narrow bins of group (halo) mass, the CLF was bimodal: the non-central (hereafter \emph{satellite}) galaxies follow a luminosity distribution close to the gamma distribution known as the \cite{Schechter:1976} function that represents well the full LF, while the central galaxies are well fit by a log-normal distribution. 
Thus, Y08 confirmed the  idea that centrals and satellites are distinct populations \citep{White.Frenk:1991,Kravtsov.etal:2004}.
These results were subsequently generalized to the SMF by \cite*{Yang.etal:2009}. 

The Yang model leads to a stellar mass - halo mass relation (SMHM) where the upper envelope increases with halo mass, and with a gap between the centrals and the satellite galaxies. This particular SMHM has been reproduced quite accurately by \cite{Cattaneo.etal:2011} in an analytical model of galaxy formation used to predict stellar masses for galaxies living in groups of given mass and at a given redshift and run on the halo merger trees extracted from a high-resolution cosmological simulation.

The idea of a gap between centrals and satellites has been around for a long time. Using N-body simulations of isolated groups, various teams recognized early on that mergers will lead to runaway growth of the most massive galaxy \citep*{Carnevali.etal:1981,Schneider.Gunn:1982,Barnes:1989}. This growth occurs independently of the merger mechanism \citep{Mamon:1987b}, whether the group evolves by direct merging between galaxies or by orbital decay via dynamical friction that causes galaxies to lose energy and angular momentum against a diffuse background (e.g., \citealp{Schneider.Gunn:1982}). In both scenarios, the growth of the central, \emph{brightest group galaxy} (BGG) occurs at the expense of the \emph{second-brightest group galaxy} (SBGG), because the merger cross-section of the SBGG is greater than that of the smaller, less luminous galaxies, and because the dynamical friction time scales as the inverse of the galaxy subhalo mass \citep{Chandrasekhar:1943a}, leading to faster orbital decay of the SBGG, hence more rapid merging with the BGG. The early N-body simulations of \cite{Mamon:1987b} showed that the gap grows as log time after a waiting period of 0.5 to 1 Gyr (see \citealt{Farhang.etal:2017}).

In the hope of finding groups that have been the sites of such rapid growth of the most massive (central) galaxy, people searched for groups with very large \emph{magnitude gaps} between their first and second-ranked members. Such studies were boosted by the discovery of diffuse X-ray sources, roughly as X-ray luminous as the most luminous groups, with a giant elliptical galaxy surrounded by low-luminosity satellites \citep{Ponman.etal:1994,Mulchaey.Zabludoff:1999}. These systems were recognized as having evolved by early galaxy mergers, which ceased long ago without replenishment of the group high luminosity galaxy population by new infalling members. \cite{Jones.etal:2003} defined \emph{fossil groups} (FGs, as coined by \citealt{Ponman.etal:1994}) to be X-ray luminous (bolometric $L_{\rm X} > 10^{42}\,h_{50}^{-2}\,\rm erg\,s^{-1}$) and with a large absolute magnitude gap: ${\cal M}_2 - {\cal M}_1 > 2$, where the SBGG is the second most luminous galaxy within a maximum projected radius of $R_{\rm max} = 0.5\,r_{200}$.
Assuming that high X-ray luminosity is a proxy for high total mass, one can study FGs with optical group catalogues, by seeking high mass groups with such large magnitude gaps.

However, without information on the group mass, only a fraction of \emph{large-gap groups} (LGGs) are in fact FGs \citep{Raouf.etal:2014}.
The origin of LGGs and FGs is still a matter of debate. 
Studies of the evolution of FGs in cosmological simulations seem to indicate
that the mass assembly histories of FGs haloes differ from those of \emph{small-gap groups} \citep[SGGs,][]{DOnghia.etal:2005, Dariush.etal:2007, Raouf.etal:2014, Farhang.etal:2017}.

The magnitude gap serves also a diagnostic of the LF.
\cite{Tremaine.Richstone:1977} derived two simple
statistics to test whether the observed magnitude gaps are consistent with
one or several cumulative luminosity functions (LFs) that 
have power-law asymptotic behavior at the faint and bright ends.
They defined the quantities 
\begin{equation}
 T_1 = \frac{\sigma(\mathcal{M}_1)}{\left\langle \mathcal{M}_2 - \mathcal{M}_1\right\rangle} \ {\rm and\ }
 T_2 = \frac{1}{\sqrt{0.677}} \frac{\sigma(\mathcal{M}_2 - \mathcal{M}_1)}{\left\langle \mathcal{M}_2 - \mathcal{M}_1\right\rangle} \ ,
 \label{Eq_TR}
\end{equation}
where $\sigma(\mathcal{M}_1)$
corresponds to the standard deviation of the absolute magnitude of the
first-ranked galaxies, while ${\left\langle \mathcal{M}_2 - \mathcal{M}_1\right\rangle}$ and ${\sigma(\mathcal{M}_2 - \mathcal{M}_1)}$ are the respective
mean and standard deviation of the magnitude gap distribution. Using Cauchy-Schwarz inequalities, \citeauthor{Tremaine.Richstone:1977} show that $T_1$ and
$T_2>1$ for groups with galaxy LFs
with power-law shapes for their cumulative forms in both their faint and
bright ends.
In group samples with $T_1$ and $T_2$ lower than unity, the
first-ranked galaxies are abnormally bright compared to the
second-ranked galaxies, unless the group samples are small ($\la 40$) or the group richnesses are
low ($\la 8$) (see table~7 of \citealp{Mamon:1987b}).

Values of $T_1$ and $T_2$ smaller than unity have been found in several
studies: \citeauthor{Tremaine.Richstone:1977} measured ${T_1 = 0.45}$ for their full cluster
sample and ${T_1 = 0.72}$ for their subsample of clusters with over 75 members.
\cite{Loh.Strauss:2006} found ${T_1 = 0.75 \pm 0.1}$ and ${T_2 = 0.86 \pm
0.1}$ in nearby rich  clusters in the Sloan Digital Sky Survey (SDSS) dominated by Luminous Red Galaxies
(LRGs).  \cite*{Lin.etal:2010} determined ${T_1 = 0.84 \pm 0.01}$ and ${T_2 =
0.77 \pm 0.01}$ for low luminosity SDSS clusters, but $T_1 = 0.70 \pm 0.1$ and
${T_2 = 0.79 \pm 0.01}$ in luminous ones.  Furthermore,
\cite{DiazGimenez.etal:2012} obtained even smaller values for a complete sample
of compact groups drawn from the 2MASS catalogue: ${T_1 = 0.51 \pm 0.06}$ and
${T_2 = 0.70 \pm 0.06}$, with slightly lower values for mock compact group
samples extracted from semi-analytical models of galaxy formation. 
These low values of $T_1$ and $T_2$ imply that  the cumulative LF can depart from power-law behaviour at the faint or bright ends.

The simple $N$-body simulations of \citet{Mamon:1987b} indicated that
galaxy mergers in groups lead to low
values of $T_1$ and $T_2$, as they tend to build the
first-ranked galaxy at the expense of the second-ranked one, as confirmed  by \cite{Smith.etal:2010} through observations.

It has been suggested that galaxy properties are not only a function of their halo mass, but also of their halo assembly history \citep{Gao.etal:2005}, which is termed \emph{galaxy assembly bias}. Perhaps the first unequivocal proof of such galaxy assembly bias was the measurement of greater dark matter concentrations of halos of red galaxies compared to blue galaxies of the same stellar or halo mass \citep{Wojtak.Mamon:2013}, given that red galaxies have older stellar populations while, at given mass, more concentrated haloes assemble earlier \citep{Wechsler.etal:2002}.

If large magnitude gaps are caused by an earlier merger history, one would then expect that the BGGs in LGGs should show older stellar populations.
In a previous study 
(\citealp*{Trevisan.etal:2017a}, hereafter Paper~I), 
we investigated whether such a difference in the evolution of LGGs is imprinted in the global properties of the stellar
content and in the star formation history of the BGGs. Through a detailed reconstruction of the 
stellar assembly of galaxies, we found that, after removing the dependence with galaxy velocity dispersion or with stellar mass, 
there is no correlation with magnitude gap of BGG ages, metallicities, and SFHs. 

This lack of variation of SFH with \deltam\ suggests that all BGGs are formed in a very similar 
way regardless of the magnitude gap. Therefore, groups with large \deltam\ might merely represent  
an evolutionary phase of galactic systems. 
Indeed, analyses of semi-analytical models indicate that most LGGs turn into regular groups by seeing the gap filled by infalling luminous galaxies \citep{Dariush.etal:2007,vonBendaBeckmann.etal:2008}.
This suggests that the ``\emph{true}'' 
SBGGs may lie further than \halfRvir\ away from the BGG \citep[e.g.][]{Gozaliasl.etal:2014b}.

It is natural to expect that the distribution of magnitude gaps should be consistent with the general LF or the CLF.
\citet{Paranjape.Sheth:2012}, 
claim that the gap distributions are consistent with
the order statistics of luminosities sampled from a single LF independent of halo mass, as long as the LF is accurate at the bright end.
This appears to be in contradiction with the 2-component CLF of Y08.
\citet{Hearin.etal:2013} analyzed both SDSS groups and mock groups to find that the richness-mass relation of groups differs between LGGs and SGGs: LGGs have higher masses (measured by their velocity dispersions) then SGGs of same richness. They therefore suggest that group masses can be made more precise once the gap is factored in.  More accurate group masses would be beneficial for cosmographic studies based on the group/cluster mass function.
Motivated by this finding, \citet{More:2012} demonstrated  that,
at fixed richness, the CLF should indeed lead to  a correlation of 
 \deltam\ with group mass. 

In this article, we wish to address three questions related to magnitude gaps of groups.
1) Is the Y08 model for the CLF correct, in particular are the satellite high-end slope as well as the ratio of satellite to central characteristic luminosities independent of group mass?
2) Is the CLF consistent with the statistics of magnitude gaps?
3) Is the richness - mass relation of groups a function of gap, allowing more accurate group mass determinations?
We will answer these questions by analyzing a doubly-complete subsample of the SDSS, suitably cleaned for edges and nearby saturated stars, using the latest Yang group catalogue.

This paper is organized as follows. In Sect.~\ref{Sec_sample}, we describe
the sample of groups and the data used in our analysis.  In Sect.~\ref{Sec_CLF}, we 
fit different CLF models, and in Sect.~\ref{Sec_TRstat}, we compute
the statistics of the bright end of these CLFs.
In Sect.~\ref{Sec_discussion}, we discuss our results and
in Sect.~\ref{Sec_summary}, we present the summary and the conclusions of our study.
We adopt WMAP3
cosmological parameters of a flat $\Lambda$CDM Universe with 
${\Omega_{\rm m} = 0.275}$, ${\Omega_{\Lambda} = 0.725}$, ${\Omega_{\rm b} = 0.046}$ and 
${H_0 = 70.2\,\hbox{km s}^{-1}\,\hbox{Mpc}^{-1}}$,
to be consistent with parameters of the group catalogue used in this study.

\section{Sample and data}
\label{Sec_sample}

The galaxy groups were selected from the updated version of the catalogue compiled by \citet{Yang.etal:2007}.\footnote{We used 
the catalogue {\tt petroB}, which is available at \url{http://gax.shao.ac.cn/data/Group.html}.} 
The new catalogue contains 473$\,$482 groups drawn from a sample of 601$\,$751 galaxies mostly from the 
Sloan Digital Sky Survey's Data Release~7 \citep[SDSS-DR7,][]{Abazajian.etal:2009}.

The radii $r_{200,{\rm m}}$, i.e, the radii of spheres that are $200$ times denser than the 
\emph{mean} density of the Universe, are derived from the $M_{200,{\rm m}}$ masses given in the \citeauthor{Yang.etal:2007}
catalogue, which are based on abundance matching with the group luminosities. 
We then calculated the virial radii (${r_{\rm vir} = r_{200,{\rm c}}}$, 
where $r_{200,{\rm c}}$ are the radii of spheres that are $200$ times denser than the 
\emph{critical}\footnote{See appendix~A in Paper~I for the conversion from quantities relative to the 
mean density to those relative to the critical density.} density of the Universe) and 
masses (${M_{\rm vir} \equiv M_{200,{\rm c}} = 200\,H^2(z)\,r_{200,{\rm c}}^3/G}$) by assuming the 
\citeauthor*{Navarro.etal:1996} (\citeyear{Navarro.etal:1996},  NFW) profile and the concentration-mass relation given by 
\cite{Dutton.Maccio:2014}. 

We used  biweight and gapper scale estimators (see \citealt{Beers.etal:1990} and references therein) to determine the group velocity dispersions, $\sigma_{\rm v}$, 
for groups with $N_{\rm vir} \geq 10$ and $N_{\rm vir} < 10$, respectively, where $N_{\rm vir}$ is the number of galaxies within $r_{\rm vir}$.
The $\sigma_{\rm v}$  values were corrected to the group rest-frame by dividing by the factor 
$(1+z_{\rm group})$ \citep{Peebles:1993}.

We selected groups that satisfy the following criteria:
\vspace{-0.5\baselineskip}
\begin{enumerate} 
\item \label{sample_z} redshifts in the  range from $0.015$ to $0.07$;

\item \label{sample_Mhalomin} masses \Mvir$\,\geq 13.0$;

\item \label{sample_Lmin} at least two member galaxies within $0.5\,r_{\rm vir}$ brighter than
  \Mabs$\,\leq -19.57$, where \Mabs\ is the k-corrected SDSS Petrosian absolute magnitude in the $r$ 
  band; 
  
\item \label{sample_gap} the magnitude gap, defined as the difference between the k-corrected SDSS
$r$-band Petrosian absolute magnitudes of the BGG and SBGG galaxies within half the  virial radius, i.e.,
\[
\Delta \mathcal{M}_{12} = \mathcal{M}_{r,2}^{\rm Petro} - \mathcal{M}_{r,1}^{\rm Petro} \ ,
\]
\noindent is smaller than $2.47\,\hbox{mag}$.
\end{enumerate}

The lower redshift limit was chosen to avoid selecting groups too close to the edge of the catalogue 
(the groups were defined using galaxies at $0.01 < z_{\rm gal} < 0.2$). 
The upper limit was optimized to obtain the largest possible number of groups with \deltam$\,\ge 2\,\hbox{mag}$, 
given the other criteria and taking into account the variation of \Mabs\ and \deltam\ limits with $z$.

To establish the \MvirNoUnit\ completeness limit, we compared the halo mass function of our sample with the
theoretical halo mass function computed using the {\sc HMFcalc} tool\footnote{\url{http://hmf.icrar.org/}} \citep*{Murray.etal:2013}. 
The adopted halo mass lower limit, \Mvir$\,\geq 13.0$, corresponds to the
value above which the difference between the observed and theoretical mass functions is smaller than~$\sim 0.1$~dex. 

We did not use the galaxy absolute magnitudes of the Yang group catalogue, but instead retrieved
the apparent  galaxy magnitudes from the SDSS-DR12 database \citep{Alam.etal:2015}. 
Since our sample is within a small range in redshift, we did not apply any evolution correction with redshift 
\citep[see e.g.][]{Blanton.etal:2003b, Yang.etal:2007}\footnote{The evolution correction to $z = 0.1$ 
determined by \citet{Blanton.etal:2003b} (${E[z]\,=\,-1.62\,[z - 0.1]}$ in the $r$-band) 
leads to corrections ranging from ${E(z\,=\,0.015) \sim -0.14}$ to $E(z = 0.07) \sim -0.05\,\hbox{mag}$ for our sample.},
and the absolute magnitudes are simply given by 
\begin{equation}
 \mathcal{M}_r^{\rm Petro} = m_r^{\rm Petro} - {\rm DM}(z) - k_{0.1}(z)  \ ,
\end{equation}
\noindent where \mr\ is the Petrosian apparent magnitude in the $r$ band corrected for Galactic extinction and ${\rm DM}(z)$ is the distance modulus.
The k-corrections, $k_{0.1}(z)$, were obtained with the {\sc kcorrect} code (version 4\_2)
of \cite{Blanton.etal:2003a}, choosing as reference the median redshift of the SDSS main galaxy sample (MGS, $z = 0.1$)\footnote{Although the median redshift of our sample is $0.05$, we kept the median redshift of the SDSS MGS as the reference for the k-corrections. 
The difference between $^{0.1}\!\mathcal{M}^{\rm Petro}_r$ and $^{0.05}\!\mathcal{M}^{\rm Petro}_r$ is $\sim 0.1$ mag, 
and has no effect on the results and conclusions of our study.}. 

We determine the 95 percent limit in  absolute magnitudes following 
the geometric approach similar to that described by \cite*{Garilli.etal:1999} and \cite{LaBarbera.etal:2010a}.  
We first determine the 95 percentile of the extinction-corrected apparent magnitude, \mr, in bins of \Mabs\ and then perform a linear fit to the 95-percentile points, 
so that the the value of  \Mabs\ where the best-fit line intersects \mr~$=~17.77$ defines the absolute magnitude
of 95 percent completeness. This leads to a 95 percent completeness limit of \Mabs~$\leq -19.57$ for our sample. 

This absolute magnitude limit in turns leads to a sample complete up to \deltam\ $= 2.47\,\hbox{mag}$, as illustrated in Fig.~1 of Paper~I. 

The criteria above lead to a sample of $2319$ groups. 

\subsection{Spectroscopic incompleteness}
\label{Sec_fiber_col}

The SDSS fibre collision limit prevents neighbouring fibres from being closer than $55''$, therefore affecting the completeness of the SDSS spectroscopy in high-density regions.
This spectroscopic incompleteness may lead to an incorrect identification of the BGGs and SBGGs. Following the approach adopted in Paper~I to address this issue, we used the SDSS photometric catalogue to identify galaxies without SDSS-DR7 spectra that could be BGGs or SBGGs. 

Concerning BGGs, we first selected all the photometric SDSS-DR12 galaxies within one virial radius from the luminosity-weighted center of each group that are brighter than the group's BGG. 
We then retrieved the spectroscopic redshifts available in SDSS-DR12 and the NASA/IPAC Extragalactic Database (NED), 
to check if these galaxies lie within the redshift range of the groups. We adopt a maximum redshift separation given by
\begin{equation}
 |z - z_{\rm group}|\ c < 2.7\,\sigma_{\rm v}\ .
 \label{Eq_dz}
\end{equation}
The factor 2.7 in equation~(\ref{Eq_dz}) provides  optimal rejection of outliers to recover the line-of-sight velocity dispersion profile expected for a single-component NFW model \citep*{Mamon.etal:2010a}. 
If the group contains a galaxy brighter than its BGG, but the galaxy has no redshift available in either SDSS-DR12 or NED, then the group is discarded.
These criteria led us to discard 228 groups that may have incorrect BGG identifications.

%
\begin{table}
\caption{Steps in cleaning the group sample for different maximum distance, \Rmax, allowed for the SBGGs.}
\centering
\tabcolsep 5pt
 \begin{tabular}{lrrrr}
 \hline
 & \multicolumn{4}{c}{$R_{\rm max}/r_{\rm vir}$}   \\
 \cline{2-5}
   \multicolumn{1}{c}{criteria}                                  &      $0.5$  &       $1.0$ &      $1.5$  &     $2.0$   \\
\hline                                          
\vspace{0.2cm}  
(1) Initial number of groups                                               &      $2319$ &      $2900$ &      $3021$ &     $3027$  \\ 
                                    
\multicolumn{2}{l}{(2) Spectroscopic incompleteness}\\
  \qquad BGG (within $1\,$\Rvir) &
      $228$ &       $271$ &       $286$ &      $289$  \\
\vspace{0.2cm}                                                                             	\qquad SBGG (within $R_{\rm max}$) &
        $192$ &       $286$ &       $367$ &      $465$  \\ 
 (3) Near edges or bright stars 
 &                                $302$ &       $349$ &       $346$ &      $353$  \\
 \\
 
 (4) {\bf Final group sample}                 & $\bm{1597}$ & $\bm{1994}$ & $\bm{2022}$ & $\bm{1920}$ \\
   \qquad LGGs &
      $152$ &       $111$ &       $93$ &      $78$  \\
\vspace{0.2cm}                                                                             	\qquad SGGs &
        $275$ &       $446$ &       $474$ &      $460$  \\ 
 
 (5) {\bf Satellites at} $\bm{R \leq R_{\rm max}}$                   & $\bm{4961}$ & $\bm{9234}$ & $\bm{11\,152}$ & $\bm{11\,095}$  \\
 \hline
 \end{tabular}

\label{Tab_samples}
\end{table}

For SBGGs, we followed a similar approach, by retrieving from the photometric catalogue all galaxies within $0.5\,r_{\rm vir}$ from the BGGs that are brighter than the SBGG of that group.
Groups are excluded if any of these galaxies have redshifts in the range specified by Eq.~(\ref{Eq_dz}), or if there is no spectroscopic redshift.
We thus discarded an additional 192 groups that may have incorrect SBGG identifications.

\subsection{Bright stars and edges of the survey}
\label{Sec_masks}

The proper identification of BGGs and SBGGs might also be affected by the presence of very bright stars lying near the line of sight that may hide or prevent accurate photometric measurement of BGGs or SBGGs.  Moreover, BGGs and/or SBGGs of groups close to the edges of the survey may lie outside the area covered by SDSS.
Therefore, we ensure that at least $95\%$ of the region within ${0.5\,r_{\rm vir}}$ from the group centres lies
within the SDSS coverage area and are not masked by bright stars.

To determine the fraction, $f_{\rm mask}$, of the area within $0.5\,r_{\rm vir}$ that lies within the SDSS masks for bright stars or that are outside the 
boundaries of the survey, we adopted the SDSS-DR7 spectroscopic angular selection function mask\footnote{We used the file {\tt sdss\_dr72safe0\_res6d.pol}, 
which can be downloaded from \url{http://space.mit.edu/~molly/mangle/download/data.html}} provided by the NYU Value-Added
Galaxy Catalog team \citep{Blanton.etal:2005} and assembled with the package {\sc Mangle 2.1} \citep{Hamilton.Tegmark:2004, Swanson.etal:2008}.
Only groups with $f_{\rm mask} < 5\%$ are selected, leading to a sample of $1597$ (out of $1899$) groups.

In summary, as displayed in Table~\ref{Tab_samples} (column $R_{\rm max}/r_{\rm m vir}=0.5$),
following the criteria 
\ref{sample_z} to \ref{sample_gap} 
listed in the beginning of Sect.~\ref{Sec_sample},
but discarding a total of $228+192=420$ groups with incomplete SDSS-DR7 spectroscopy (see Sect.~\ref{Sec_fiber_col}), 
and $302$ additional groups that are in the edge of the survey and/or affected by the presence of bright stars, we 
obtain our final sample of $1597$ groups, among which $152$ have \deltam$\,> 2\,\hbox{mag}$.

\subsection{Samples with  \Rmax$\,> 0.5\,$\Rvir}
\label{Sec_samples_r}

As mentioned in Sect.~\ref{Sec_intro}, the value of \deltam\ might vary with the maximum distance allowed for obtaining the SBGG.
To investigate how the gap varies with \Rmax,  we select samples with \deltam\ defined within $0.5$, $1.0$, $1.5$, and $2.0\,r_{\rm vir}$, 
following the same criteria listed in Sect.~\ref{Sec_sample}. 
Groups with incomplete SDSS-DR7 spectroscopy, near the edges of the survey or with bright stars nearby
were excluded following the same approach described in Sections~\ref{Sec_fiber_col} and \ref{Sec_masks}.

The group samples are summarized in Table~\ref{Tab_samples}. 
The samples defined within \halfRvir\ and in one virial radius have $1425$ groups in common, and $1237$ groups are present in all four samples.

\section{Conditional luminosity functions}
\label{Sec_CLF}

The galaxy CLF can be estimated by directly counting galaxies in
groups of
 our doubly complete sample of galaxies.
The absolute Petrosian magnitudes are converted to luminosities using
the solar absolute magnitude in the $r$-band, redshifted  to $z = 0.1$, 
$\mathcal{M}_{r \odot} = 4.76$ \citep{Blanton.etal:2003b}.

As in Y08, we assume that the total CLF of haloes with $M_{\rm vir} = M$ is the sum of the CLFs of the central and the satellite galaxies:
\begin{equation}
 \Phi(L|M) =  \dv{N}{\log L} = \Phi_{\rm c}(L|M) + \Phi_{\rm s}(L|M) \ .
 \label{Eq_phi_tot}
\end{equation}
The central galaxies follow a lognormal distribution,
\begin{equation}
 \Phi_{\rm c}(L|M) = \frac{1}{\sqrt{2 \pi}\ \sigma_{\rm c}} \exp \left[ - \frac{(\log L - \log L_{\rm c})^2}{2\ \sigma_{\rm c}^2} \right]\ , 
 \label{Eq_phi_cen}
\end{equation}
\noindent while the satellite luminosities are distributed according to  the probability distribution function
\begin{eqnarray}
 \Phi_{\rm s}(L|M) &=& {\beta \over \Gamma\left[(\alpha\!+\! 1)/\beta,\left(L_{\rm min}/L_{\rm s})^{\beta}\right)\right]} \nonumber \\
&\mbox{}& \quad \times \left( \frac{L}{L_{\rm s}} \right)^{\alpha + 1} 
\exp \left[- \left( \frac{L}{L_{\rm s}} \right)^{\beta}\right]\ ,
 \label{Eq_phi_sat}
\end{eqnarray}
\noindent where $\Gamma(a, x) = \int_x^\infty t^{a-1} \exp(-t)\,{\rm d}t$ is the upper incomplete Gamma function.
%
Equation~(\ref{Eq_phi_sat}) corresponds to the \citet{Schechter:1976} LF when $\beta = 1$, while Y08 adopted $\beta = 2$ and $L_{\rm s} = L_{\rm c}/10^{0.25}$.
Hereafter, a galaxy is referred as central (satellite) if its luminosity is drawn from the CLF specific to the centrals (satellites, eqs.~\ref{Eq_phi_cen} and \ref{Eq_phi_sat}).

Instead of fitting the CLFs in bins of halo mass, we adopt a different approach. We assume that the parameters $\log L_{\rm c}$, $\sigma_{\rm c}$, $\log L_{\rm s}$, 
$\alpha$, and $\beta$ vary linearly with \MvirNoUnit, and determined the best-fit parameters through \emph{maximum likelihood estimation} (MLE).
We assume that the brightest galaxy is always the central galaxy and we fit $\Phi_{\rm c}(L|M)$ and $\Phi_{\rm s}(L|M)$ separately.
The probability of observing a galaxy with luminosity $L_i$ in a halo with mass $M_i$ is given by
\begin{equation}
  p(L_i | M_i) = \Phi(L_i|M_i) \, p(M_i) \ ,\\
 \label{Eq_p_Li}
\end{equation}
\noindent where $\Phi = \Phi_{\rm c}$ or $\Phi_{\rm s}$ for central and satellite galaxies, respectively, and the probability of observing a halo with mass $M_i$, $p(M_i)$, is given by the halo mass function.  We estimate $p(M_i)$ by fitting a second order polynomial in log-log to the distribution of halo masses from the  Yang catalogue (converted from overdensities of 200 times the mean density of the Universe to 200 times the critical density of the Universe). 

Given the observed luminosities, $\bm{\hbox{L}_{\rm c}}$, of a sample with $N_{\rm c}$ central galaxies, from eqs.~(\ref{Eq_phi_cen}) and (\ref{Eq_p_Li}) we obtain that the likelihood of a CLF described by the parameters ${\bm{\theta} = \left\{\log L_{\rm c}, \sigma_{\rm c}\right\}}$ is given by
\begin{eqnarray}
 \!\!\!\! - \ln \mathcal{L}_{\rm c}(\bm{\theta} | \bm{\hbox{L}_{\rm c}})
 &\!\!\!\!=\!\!\!\!& -\ln \prod_{i = 1}^{N_{\rm c}} \Phi_{\rm c}(L_i|M_i) \, p(M_i) \nonumber \\ 
  &\!\!\!\!=\!\!\!\!& \sum_{i = 1}^{N_{\rm c}} \ln (2 \pi\,\sigma_{\rm c}) 
  + \sum_{i = 1}^{N_{\rm c}} \frac{(\log L_i - \log L_{\rm c})^2}{2\ \sigma_{\rm c}^2} 
 \nonumber \\
 &\mbox{}& \qquad
  - \sum_{i = 1}^{N_{\rm c}} \ln p(M_i) \ ,
 \label{Eq_like_cen}
\end{eqnarray}
where $\sigma_{\rm c}$ and $L_{\rm c}$ are functions of the halo mass $M_i$.
Similarly, for a sample of $N_{\rm s}$ satellite galaxies with luminosities $\bm{\hbox{L}_{\rm s}}$, 
the likelihood of a CLF described by the parameters 
${\bm{\theta} = \left\{\log L_{\rm s}, \alpha, \beta\right\}}$ is given by
\begin{eqnarray}
 - \ln \mathcal{L}_{\rm s}(\bm{\theta} | \bm{\hbox{L}_{\rm s}}) 
 &\!\!\!\!=\!\!\!\!&  -\ln \prod_{i = 1}^{N_{\rm s}} \Phi_{\rm s}(L_i|M_i) \, p(M_i)  \nonumber \\ 
  &\!\!\!\!=\!\!\!\!& \sum_{i=1}^{N_{\rm s}}
\ln\!\left\{ \frac{1}{\beta}\,
\Gamma \left[{\alpha+1\over \beta}, \left({L_{\rm min}\over L_{\rm s}}\right)^\beta \right] \right\} \nonumber \\
&\!\!\!\!\mbox{}\!\!\!\!& \quad 
 - \sum_{i=1}^{N_{\rm s}}(\alpha+1)\,\ln  \left( \frac{L_i}{L_{\rm s}} \right)
+ \sum_{i = 1}^{N_{\rm s}} \left( \frac{L_i}{L_{\rm s}} \right)^{\beta} 
 \nonumber \\
  &\!\!\!\!\mbox{}\!\!\!\!& \quad 
  - \sum_{i = 1}^{N_{\rm s}} \ln p(M_i) \ ,
 \label{Eq_like_sat}
\end{eqnarray}
where $\alpha$, $\beta$, and $L_{\rm s}$ are functions of halo mass $M_i$.

We performed the MLE, first for the centrals, using Eq.~(\ref{Eq_like_cen}), and then for the satellites, using Eq.~(\ref{Eq_like_sat}). 
We performed the minimisation using the function {\tt mle2} from the R package {\tt bbmle} \citep{bbmle:2017},
 adopting the method L-BFGS-B \citep{Byrd.etal:1995}, which allows us to specify upper and lower bounds for each variable.
The uncertainties in the parameters were estimated by bootstrapping the sample 200 times.

%
\begin{figure}
\centering
\includegraphics[width=\hsize]{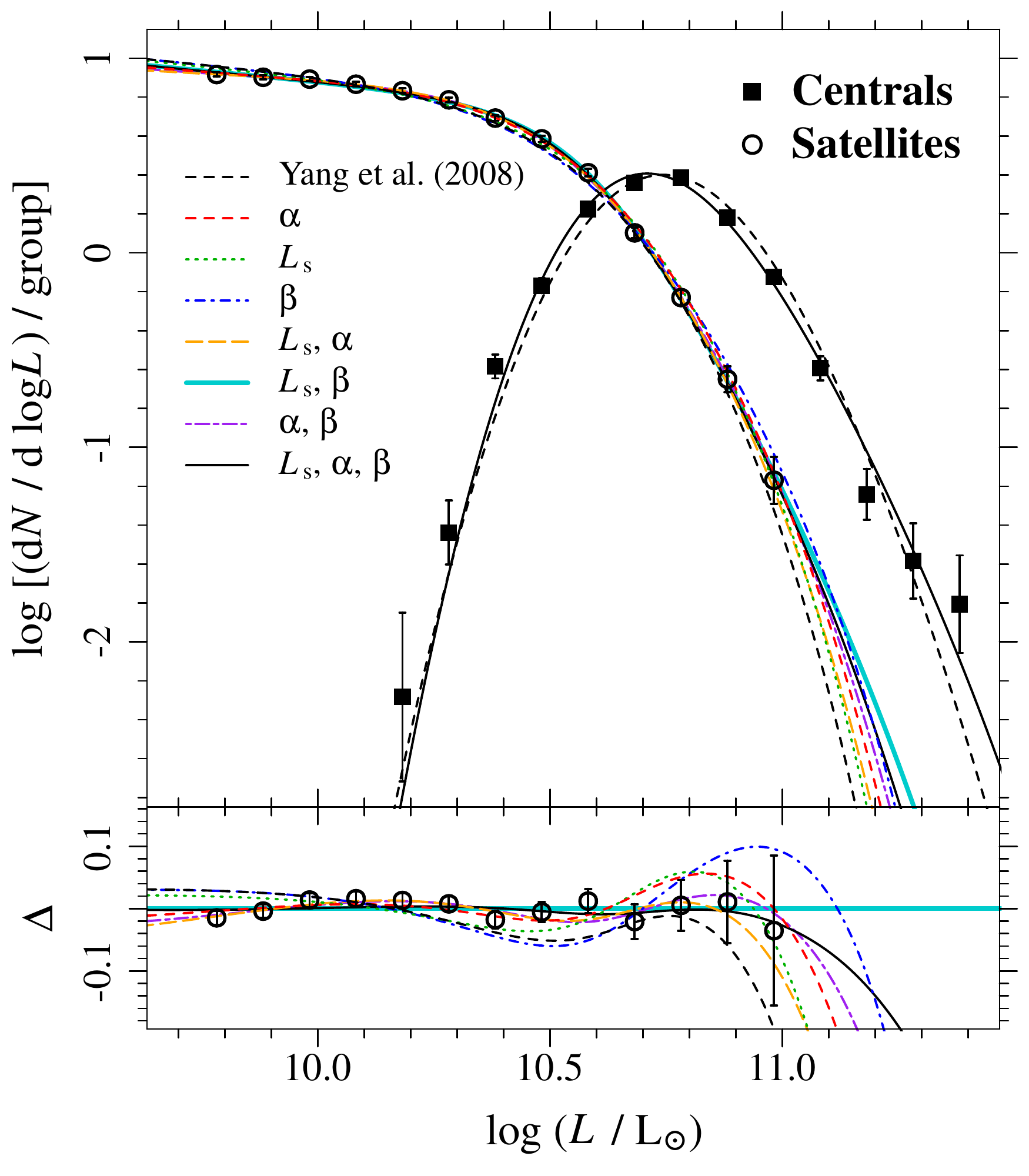}
\caption{Total luminosity functions of central and satellite galaxies, computed as the best-fit CLFs integrated over the halo masses of the groups in our sample, i.e, ${\Phi(L) = \sum_{i = 1}^{N_{\rm groups}} \Phi(L|M_i) / N_{\rm groups}}$.
For the satellites, we show the best-fit models with 2 free parameters, i.e., when we fit the linear relation with \MvirNoUnit\ of 
only one of the parameters $\alpha$ (\emph{red dashed}), $\log L_{\rm s}$ (\emph{green dotted}), 
or $\beta$ (\emph{blue dot-dashed lines}).
The \emph{orange long-dashed}, \emph{cyan solid}, and \emph{purple short-long dashed lines} indicate the models with 4 degrees of freedom, and the model shown as the \emph{black solid line} has 6 free parameters.
The data is shown as \emph{open circles} (satellite galaxies) and 
\emph{filled squares} (central galaxies). 
The luminosity functions by \citet{Yang.etal:2008} are indicated as \emph{black dashed lines}.
The lower panel shows the residuals of the satellite LF relative to the `$L_{\rm s}, \beta$' model.
}  
\label{Fig_fitCLF_sbggClean}
\end{figure}

%
\begin{figure*}
\centering
\begin{tabular}{cc}
\vspace{1mm}
 \Large{\hspace{10mm} ${13.00 \leq \log (M_{\rm vir} / M_{\odot}) < 13.35}$} &
 \Large{\hspace{10mm} ${13.35 \leq \log (M_{\rm vir} / M_{\odot}) < 13.72}$} \\
 \includegraphics[width=0.45\hsize,page=1]{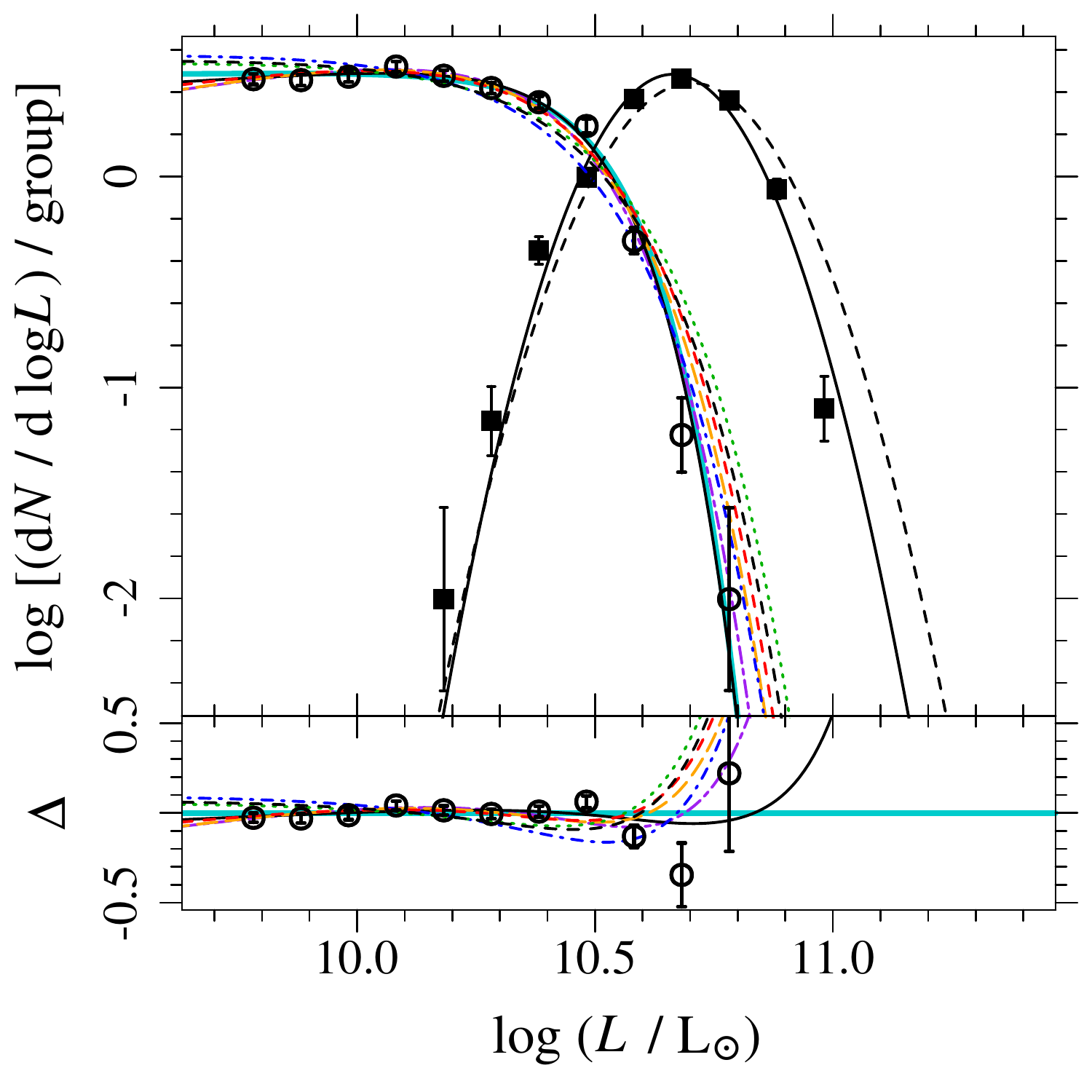}  &
 \includegraphics[width=0.45\hsize,page=2]{figs/satLF_binsMhalo_sat0-2rvir_clean_mask_BGG_SBGG_dM12.pdf}  \\
 \Large{\hspace{10mm} ${13.72 \leq \log (M_{\rm vir} / M_{\odot}) < 14.12}$}   &
 \Large{\hspace{10mm} ${14.12 \leq \log (M_{\rm vir} / M_{\odot}) \leq 15.07}$} \\
 \includegraphics[width=0.45\hsize,page=3]{figs/satLF_binsMhalo_sat0-2rvir_clean_mask_BGG_SBGG_dM12.pdf}  &
 \includegraphics[width=0.45\hsize,page=4]{figs/satLF_binsMhalo_sat0-2rvir_clean_mask_BGG_SBGG_dM12.pdf}  \\
\end{tabular}

\caption{Conditional luminosity functions in four bins of group (halo) mass. Same notations as in Fig.~\ref{Fig_fitCLF_sbggClean}.
Each bin contains the same number of groups (480 groups).
The  log luminosity axes are the same for the four panels. 
} 
\label{Fig_CLF_bins}
\end{figure*}

\subsection{Central galaxies}
\label{Sec_CLF_central}

Figure~\ref{Fig_fitCLF_sbggClean} displays the luminosity function of central galaxies (solid squares), 
computed as the best-fit CLFs averaged over the halo masses of the groups in our sample, 
i.e, ${\Phi(L) = \sum_{i = 1}^{N_{\rm groups}} \Phi(L|M_i) / N_{\rm groups}}$.
Our fits are very close to those of Y08.

\begin{figure}
\centering
\includegraphics[width=0.9\hsize]{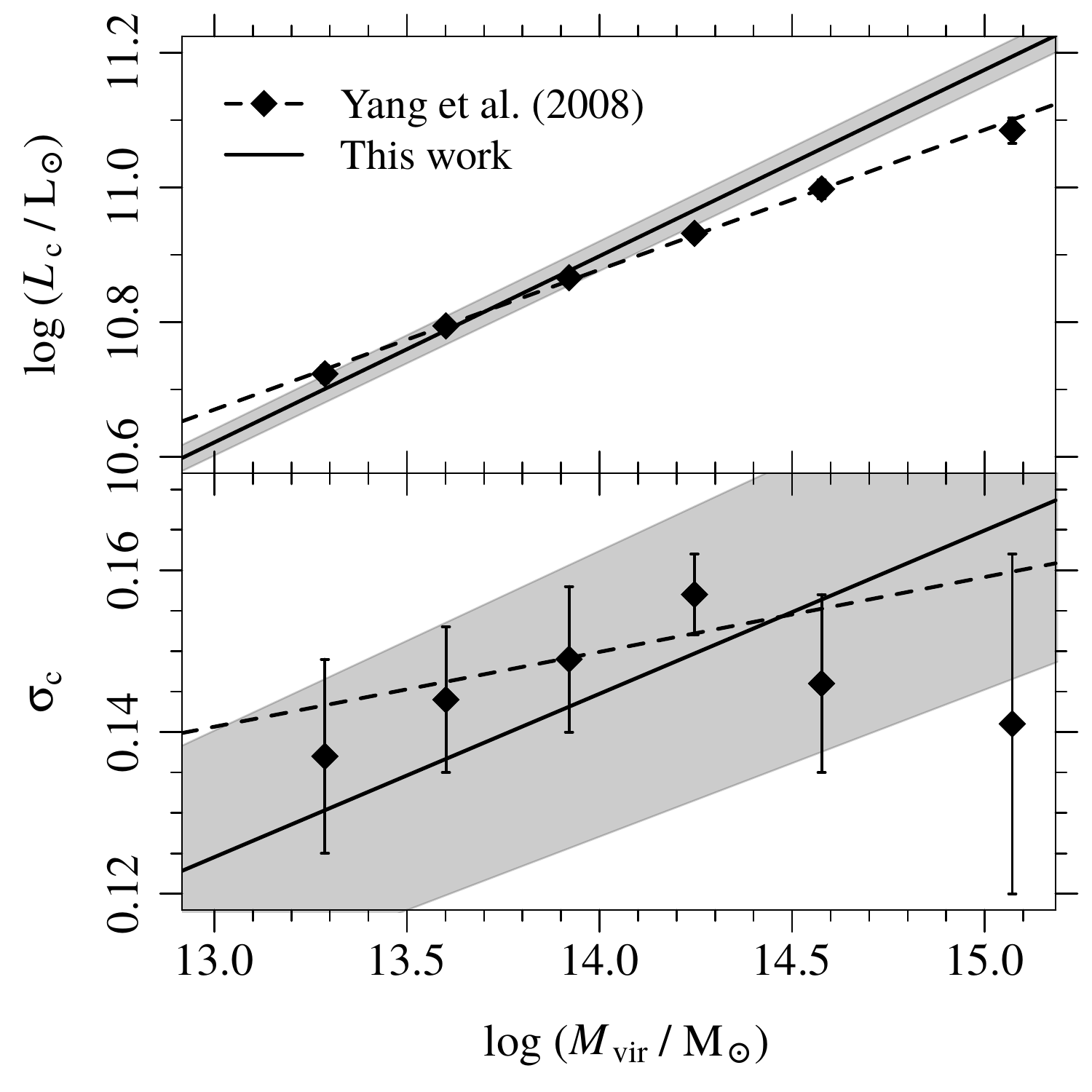}
\caption{Relation between $L_{\rm c}$ (\emph{top}) and $\sigma_{\rm c}$ (\emph{bottom}) with the group halo mass. Our best-fit relations and errors are indicated by 
the \emph{solid lines} and \emph{shaded areas}. The \emph{black squares} represents the results from \citet{Yang.etal:2008} in bins of $M_{\rm vir}$, 
and the \emph{dashed lines} show the best linear fits to their values.} 
\label{Fig_model_central}
\end{figure}

Figure~\ref{Fig_CLF_bins} shows our best-fit model in bins of halo mass.
The relations between $\log L_{\rm c}$, $\sigma_{\rm c}$ and halo mass are given by 
\begin{eqnarray}
\!\!\!\!\!\!\!\log\left({L_{\rm c} \over {\rm L}_{\odot} }\right) 
&\!\!\!\!=\!\!\!\!& (10.90\pm0.01) + (0.28\pm0.01) 
\,\log\left( {M_{\rm vir} \over {10^{14}\,\rm M}_{\odot}} \right)  
\nonumber \\
\!\!\!\!\!\!\!\!{\sigma_{\rm c} \over {\rm L}_{\odot}} 
&\!\!\!\!=\!\!\!\!& (0.15\pm0.01) + (0.02\pm0.01) 
\,\log\left({M_{\rm vir} \over {10^{14}\,\rm M}_{\odot}} \right)\,.
\end{eqnarray}
In comparison with Y08, our centrals have lower luminosity at low group mass and higher luminosity at high group mass. 
This is illustrated 
in Figure~\ref{Fig_model_central}, where we compare our CLF fits and to the CLFs of Y08 and their fits.
Their results are given in bins of $M_{180,m}$, i.e., the mass within spheres that are $180$ times denser than the 
mean density of the Universe, which we converted to $M_{200,c}$. Moreover,
their luminosities are corrected for evolution. 
After estimating it from our sample, we subtracted the evolutionary correction from the
Y08 $\log L_{\rm c}$ values and fitted the relations with \MvirNoUnit. We find 
${\log (L_{\rm c} / \hbox{L}_{\odot}) = 10.88 + 0.21 \log [M_{\rm vir} / ({10^{14}\,\rm M}_{\odot})]}$ 
and ${\sigma_{\rm c} / \hbox{L}_{\odot} = 0.15 + 0.01 \log [M_{\rm vir} / ({10^{14}\,\rm M}_{\odot})]}$, 
with errors in the coefficients of $0.01$.

Our relation between $\log L_{\rm c}$ and halo mass is indeed steeper than that of Y08, and the central galaxies in haloes with \Mvir$\,\sim 15$ are $25\%$ ($0.1\,$dex) more luminous than predicted by the Y08 CLFs. 
It is not clear what is the source of this discrepancy, but it might be due to differences in the definitions of the samples:
Y08 include galaxies up to $z = 0.2$, while our sample contains only groups at $z < 0.07$. Besides, their analysis was based on SDSS-DR4, while ours is defined from DR7, and we use DR12 photometric measurements, 
which include a improved background subtraction in very luminous galaxies as the centrals in the most massive groups. 
We discuss this further in Sect.~\ref{Sec_CLF_discussion}.

%

\subsection{Satellite luminosity functions}
\label{Sec_CLF_sat}

\begin{figure}
\centering
\includegraphics[width=\hsize]{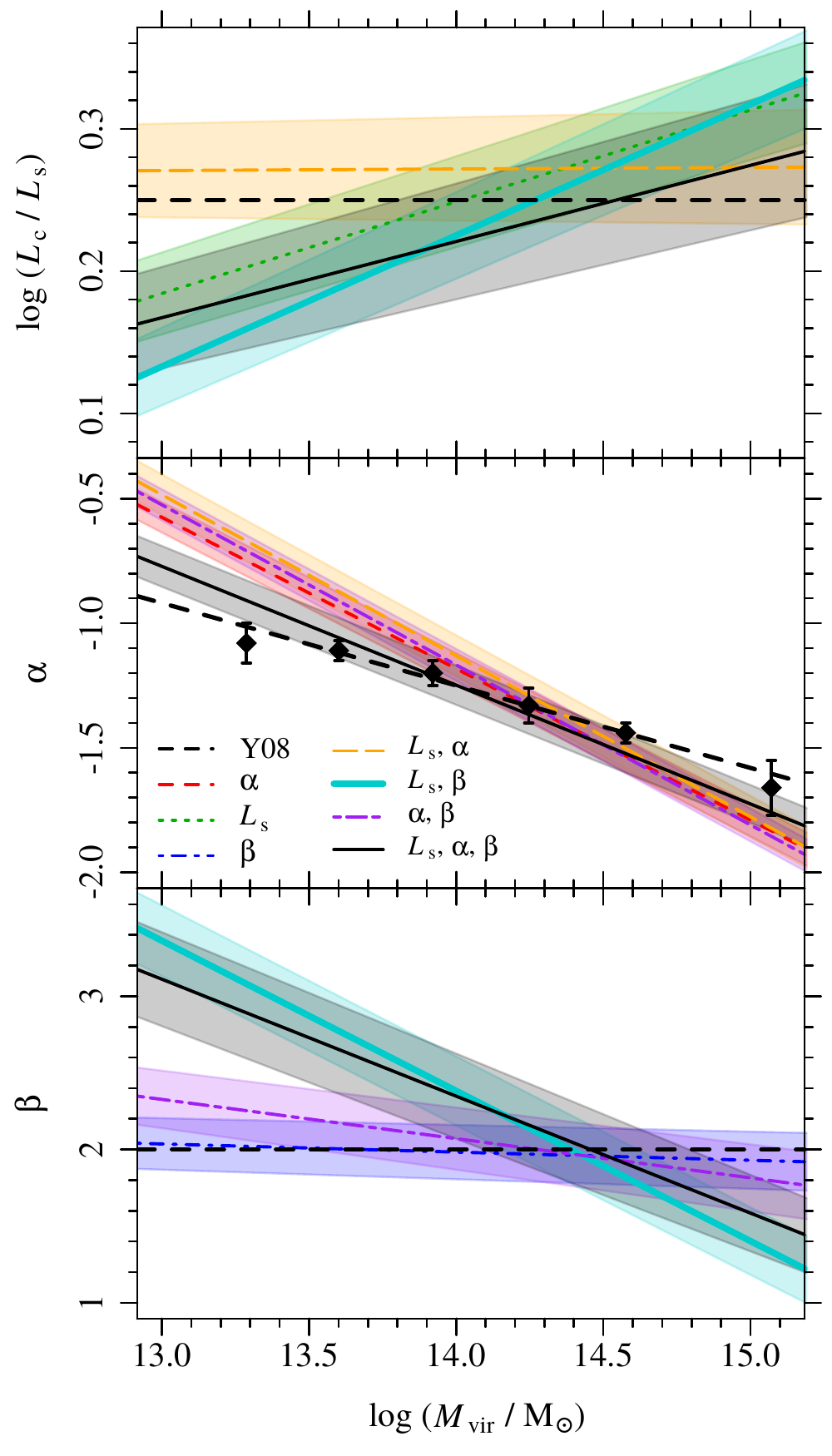}
\caption{Relations with group halo mass of $\log (L_{\rm c} / L_{\rm s})$ (\emph{upper}), $\alpha$ (\emph{middle}), 
$\beta$ (\emph{lower panel}) for CLFs with different degrees of freedom.} 
\label{Fig_model_sat}
\end{figure}

Following the same approach adopted for the central galaxies, we performed MLE to obtain the parameters of the satellite CLFs. 
We compared models with different degrees of freedom. 
We assume that the three parameters $L_{\rm s}$ (characteristic luminosity), $\alpha$ (faint-end slope), and $\beta$ (bright-end shape) vary linearly with 
\MvirNoUnit, but allow ourselves to fix instead one or several of these three parameters to their Y08 values (thus independent of $M_{\rm vir}$). The models are listed in Table~\ref{Tab_models}, where we indicate which of these relations are fitted in each model. In  models 3, 4, and 6, $\alpha$ is fixed to the Y08 relation. 
The Y08 linear relations with \MvirNoUnit\ were obtained by fitting their values of $\alpha$ in bins of halo mass (middle panel in Fig.~\ref{Fig_model_sat}). 
In models 2, 4, and 7, we fix $\log L_{\rm s} = \log L_{\rm c} - 0.25$ (as Y08), and in models  2, 3, and 5, $\beta = 2$ (as Y08).

Figure~\ref{Fig_fitCLF_sbggClean} displays the best-fit CLFs  integrated over the halo masses of the groups in our sample 
and Figure~\ref{Fig_CLF_bins} shows our best-fit models in bins of halo mass.
The relations with halo mass of $\log (L_{\rm c}/L_{\rm s})$, $\alpha$, and $\beta$ are presented in 
Figure~\ref{Fig_model_sat}. We find a positive correlation between $\log (L_{\rm c}/L_{\rm s})$ and \MvirNoUnit\ for all the models
where we allow $L_{\rm s}$ to vary, except for model 5, for which $\log (L_{\rm c}/L_{\rm s}) \sim 0.27$ (basically constant).
As in Y08, we also find that $\alpha$ decreases with halo mass; however, our best-fit relations are steeper than those of  Y08. Finally, $\beta$ is roughly constant for models 4 and 5, but in models 6 and 8, it varies from $\beta \sim 3.4$ 
to $\sim 1.4$ between \Mvir$\,= 13$ to $15$. 

\begin{table}
\centering
\caption{Comparison between satellite CLF models. {\bf (1)} Free parameters of the model; 
{\bf (2)} $p$-values of a Kolmogorov-Smirnov and {\bf (3)} Anderson-Darling tests;
{\bf (4)} $\Delta$BIC and {\bf (5)} $\Delta$AIC relative to the model with the lowest BIC (model 6) 
and AIC (model 8) values. Since model 1 has no free parameters, the values of BIC and AIC
correspond to $-2 \ln \mathcal{L}_{\rm s}$.}
 \begin{tabular}{lccrr}
 \hline                                               
 \multicolumn{1}{c}{Model}               &          \multicolumn{2}{c}{$p$-values} & $\Delta$BIC & $\Delta$AIC \\
                                         &            KS-test &            AD-test &             &             \\
 \multicolumn{1}{c}{(1)}          & \multicolumn{1}{c}{(2)} & \multicolumn{1}{c}{(3)} & \multicolumn{1}{c}{(4)}  & \multicolumn{1}{c}{(5)} \\
 \hline
 1. Yang et al. (2008)                   & $2 \times 10^{-4}$ & $9 \times 10^{-6}$ & {\it 64.7}  & {\it 100.3} \\
 2. $\alpha$                             & $6 \times 10^{-1}$ & $6 \times 10^{-1}$ &        4.7  &      25.7 \\
 3. $L_{\rm s}$                          & $2 \times 10^{-2}$ & $7 \times 10^{-3}$ &       66.4  &      87.4 \\
 4. $\beta$                              & $7 \times 10^{-5}$ & $4 \times 10^{-6}$ &      123.1  &     144.1 \\
 5. $L_{\rm s}$, $\alpha$                & $7 \times 10^{-1}$ & $5 \times 10^{-1}$ &       16.0  &      22.3 \\
 6. $L_{\rm s}$, $\beta$                 & $1 \times 10^{-1}$ & $4 \times 10^{-2}$ & $\bm{0.0}$  &       6.4 \\
 7. $\alpha$, $\beta$                    & $6 \times 10^{-1}$ & $6 \times 10^{-1}$ &        5.6  &      12.0 \\
 8. $L_{\rm s}$, $\alpha$, $\beta$       & $2 \times 10^{-1}$ & $7 \times 10^{-2}$ &        8.3  & $\bm{0.0}$\\
 \hline
\end{tabular}
\label{Tab_models}
\end{table}

We used the Kolmogorov-Smirnov (KS) and Anderson-Darling (AD) tests to check the goodness of fit of each model. 
The resulting $p$-values are shown in Table~\ref{Tab_models}. 
The $p$-values of the AD tests are typically smaller than those of the KS tests, since the former is more sensitive to the tails of the distributions. 
Both tests indicate that models 1 (Y08), 3 (free $\log L_{\rm s}$), and 4 (free $\beta$) do not provide a good description of the data ($p < 0.05$).

We also computed the \emph{Bayesian information criterion} (BIC) and the \emph{Akaike information criterion} (AIC). These criteria are a 
measure of the relative quality of statistical models for a given set of data, and the model with the lowest BIC/AIC is preferred. 
There is strong evidence against a model when $\Delta$BIC$\, > 6$, and it is decisive when $\Delta$BIC$\, > 10$ \citep{Raftery:1995, Kass.Raftery:1995}.

Table~\ref{Tab_models} shows the values of BIC and AIC relative to the best models according to each of these criteria (i.e., models 6 and 8, respectively).
There is a disagreement between these two criteria: BIC indicates that model 6 is the best model, while model 8 is the preferred one according to AIC
(BIC penalises model complexity more heavily, while AIC leads to more complex models).

Therefore, models 1, 3, and 4 can be clearly rejected based on the $p$-values, $\Delta$BIC, and $\Delta$AIC.
Models 2, 5, and 7 can be discarded according to $\Delta$BIC or $\Delta$AIC (or both).
Model 8 corresponds to the best model based on $\Delta$AIC, but there is strong evidence against it based on $\Delta$BIC. 
Finally, model 6 is the best model based on $\Delta$BIC, and there is only marginal rejection based on the AD test and $\Delta$AIC.
Therefore, we adopt model 6 as our standard model throughout the paper.
Moreover, model 6 has the advantage  of adopting  the value of $\alpha$ from Y08 who probe the CLFs to fainter luminosities with their flux-limited sample.
However, as we discuss later, the main conclusions of our study do not depend on the particular choice of the model. 

%

%

%
\begin{table*}
\centering
\caption{Parameters of the satellite CLF models. {\bf (1)} Free parameters of the model;
{\bf (2)}, {\bf (4)}, and {\bf (6)} Values of the parameters $L_{\rm s}$, $\alpha$, and $\beta$ for $M_{\rm vir} = 10^{14} {\rm M}_{\odot}$, respectively.
{\bf (3)}, {\bf (5)}, and {\bf (7)} Slopes of the linear relations with \MvirNoUnit. Our preferred model is indicated in bold.}
\tabcolsep 4pt
 \begin{tabular}{lcccccc}
 \hline                                               
 \multicolumn{1}{c}{Model}            & $\log (L_{\rm s}/{\rm L}_{\odot}) $ & ${\rm d}\log L_{\rm s} / {\rm d}\log M_{\rm vir}$  
                                      & $\alpha$  & ${\rm d}\alpha / {\rm d}\log M_{\rm vir}$ 
                                      & $\beta$   & ${\rm d}\beta / {\rm d}\log M_{\rm vir}$  \\
 
                                      & [$M_{\rm vir} = 10^{14} {\rm M}_{\odot}$] & &[$M_{\rm vir} = 10^{14} {\rm M}_{\odot}$]  & & [$M_{\rm vir} = 10^{14} {\rm M}_{\odot}$] \\
 \multicolumn{1}{c}{(1)}              & (2) & (3) & (4) & (5) & (6) & (7)  \\
 \hline
 1. Yang et al. (2008)                &    $ 10.63 \pm 0.01 $  &     $ 0.21 \pm 0.01 $ &    $ -1.25 \pm 0.01 $ &    $ -0.33 \pm 0.02 $ &             $2.00\ \ \ \ \ \ \ \ $ &             $0.00\ \ \ \ \ \ \,$ \\
 2. $\alpha$                          &              $10.65\ \ \ \ \ \ \ \ $  &              $ 0.28\ \ \ \ \ \ \ \ $ &      $-1.18 \pm 0.04$ &      $-0.61 \pm 0.03$ &             $ 2.00\ \ \ \ \ \ \ \ $ &             $  0.00\ \ \ \ \ \ \,$ \\
 3. $L_{\rm s}$                       &      $10.65 \pm 0.01$  &       $0.21 \pm 0.01$ &              $ -1.25\ \ \ \ \ \ \ \ $&             $ -0.33\ \ \ \ \ \ \ \ $ &             $ 2.00\ \ \ \ \ \ \ \ $ &             $  0.00\ \ \ \ \ \ \,$ \\
 4. $\beta$                           &              $ 10.65\ \ \ \ \ \ \ \ $  &              $ 0.28\ \ \ \ \ \ \ \ $ &              $-1.25\ \ \ \ \ \ \ \ $ &             $-0.33\ \ \ \ \ \ \ \ $ &      $1.98 \pm 0.11$ & $-0.05 \pm 0.07$ \\
 5. $L_{\rm s}$, $\alpha$             &      $10.63 \pm 0.02$  &       $0.28 \pm 0.01$ &      $-1.13 \pm 0.05$ &      $-0.65 \pm 0.03$ &             $ 2.00\ \ \ \ \ \ \ \ $ &             $  0.00\ \ \ \ \ \ \,$  \\
{\bf 6.} $\bm{L_{\rm s}}$, $\bm{\beta}$& $\,\bm{10.67 \pm 0.01}$  &  $\bm{0.19 \pm 0.01}$ &          $\bm{-1.25}\ \ \ \ \ \ \ \ $ &        $\bm{ -0.33}\ \ \ \ \ \ \ \ $ & $\bm{2.38 \pm 0.14}$ & $\bm{-0.98 \pm 0.09}$ \\
 7. $\alpha$, $\beta$                 &              $ 10.65\ \ \ \ \ \ \ \ $  &              $ 0.28\ \ \ \ \ \ \ \ $ &      $-1.17 \pm 0.04$ &      $-0.64 \pm 0.03$ &      $2.07 \pm 0.12$ & $-0.26 \pm 0.08$ \\
 8. $L_{\rm s}$, $\alpha$, $\beta$    &      $10.68 \pm 0.02$  &       $0.22 \pm 0.01$ &      $-1.25 \pm 0.05$ &      $-0.48 \pm 0.03$ &      $2.35 \pm 0.20$ & $-0.76 \pm 0.10$ \\
 \hline 
\end{tabular}
\label{Tab_models_pars}
\end{table*}

\section{Magnitude gap statistics from the conditional luminosity functions}
\label{Sec_TRstat}

\subsection{Gap statistics, predictions from the conditional luminosity function, and previous results}
\label{Sec_gapstats_1}
%
\begin{figure}
\centering
\begin{tabular}{c}
 \Large{$\bm{R_{\rm max} = 0.5\,r_{\rm vir}}$} \\ 
 \includegraphics[width=0.99\hsize]{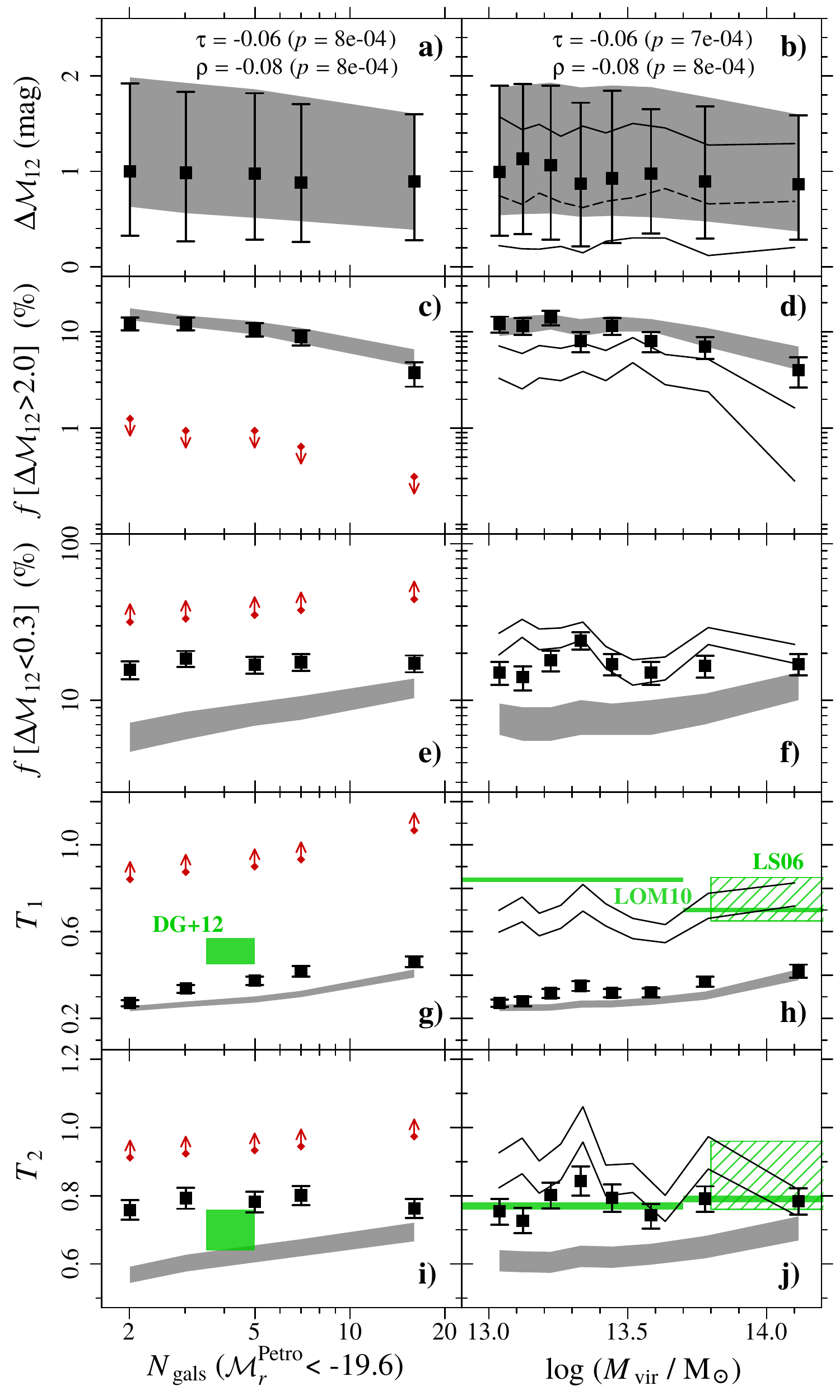} 
\end{tabular}

\caption{Statistics of the magnitude gap as a function of group richness (\emph{left}) and halo mass (\emph{right}):
magnitude gap ({\bf a, b}); fractions of groups with \deltam$\, > 2.0$ ({\bf c, d}) and $< 0.3$ mag 
({\bf e, f}); $T_1$ ({\bf g, h}) and $T_2$ ({\bf i, j}).
In panels {\bf a, b}, the \emph{black squares} and errors bars
represent the median, the $16^{\rm th}$ and the $84^{\rm th}$ percentiles of the \deltam~distribution
in bins of $N_{\rm gals}$ and \MvirNoUnit. The Kendall ($\tau$) and Spearman
($\rho$) rank correlation coefficients, as well as the $p$-values, are
indicated in each panel. In panels {\bf c--f}, the error bars
represent the binomial errors. In panels {\bf c, d}, the errors were
estimated by bootstraping the sample $1000$ times, and errors bars indicate
the $16\%$ and $84\%$ percentiles. The \emph{shaded gray areas} represent
the $16\%$ and $84\%$ percentiles of the results from sampling the CLF of model~6 in Table~\ref{Tab_models}.
The {\emph{red symbols}} in panels {\bf g, i} indicate the 5 ({\emph{arrows pointing upward}}) or 
95 ({\emph{arrows pointing downward}}) percentiles of the values expected for the \citet{Schechter:1976} LF with $\alpha = -1.05$.
We also show the $T_1$ and $T_2$ values from studies by \citet[][DG+12]{DiazGimenez.etal:2012}, 
\citet[][LS06]{Loh.Strauss:2006}, and \citet[][LOM10]{Lin.etal:2010}. 
In the panels showing the statistics versus halo mass, the solid black lines indicate the equivalent $\pm  1\,\sigma$ range of values estimated from the semi-analytical model of \citet{Henriques.etal:2015}.
} 
\label{Fig_dMag_N_Mh}
\end{figure}

We now test the CLF using the statistics of  the magnitude gap $\Delta {\cal M}_{12}={\cal M}_2-{\cal M}_1$.
Figure~\ref{Fig_dMag_N_Mh} illustrates five  statistics of the gap as a function of group richness (number of galaxies, $N_{\rm gals}$,
with \Mabs$\,\leq -19.57$, \emph{left}) and halo mass $M_{\rm vir}$ (\emph{right}).
We can see that \deltam\ slightly decreases with $N_{\rm gals}$ and \MvirNoUnit (Figs.~\ref{Fig_dMag_N_Mh}a,b). The median value of the magnitude gap varies from \deltam$\,\sim 1.0\,\hbox{mag}$ in groups with $N_{\rm gals} = 2$ to \deltam$\,\sim 0.9\,\hbox{mag}$ in groups with high multiplicity ($N_{\rm gals} \gtrsim 10$). A similar variation is observed in Figure~\ref{Fig_dMag_N_Mh}b: \deltam~decreases from $\sim 1.0\,\hbox{mag}$ in low mass haloes (\Mvir$\,< 13.1$) to \deltam$\,\sim 0.9\,\hbox{mag}$ in haloes with \Mvir$\,\gtrsim 14$. The Spearman and Kendall correlation tests confirm the correlation of gap with $N_{\rm gals}$ ($p = 8 \times 10^{-4}$) and \MvirNoUnit\ ($p = 7-8 \times 10^{-4}$).

The fraction of LGGs decreases with $N_{\rm gals}$ and halo mass (panels c,d in Fig.~\ref{Fig_dMag_N_Mh}). 
Around $12\%$ of low-multiplicity groups have large magnitude gaps, and this fraction decreases to $\sim 4\%$ among rich groups. 
The fraction of SGGs is roughly constant and varies between $10$ and $20\%$.

We applied the \cite{Tremaine.Richstone:1977} statistics (see
Sect.~\ref{Sec_intro} and Eq.~\ref{Eq_TR}) to test if the observed magnitude
gaps are consistent with a cumulative LF that has  power law
behaviour at the faint and bright ends. 
The values of $T_1$ and $T_2$ are presented in Figs.~\ref{Fig_dMag_N_Mh}g to \ref{Fig_dMag_N_Mh}j. 
They were computed in the bins of $N_{\rm gals}$ and \MvirNoUnit. We find very low values of $T_1$ ($\la 0.5$) and $T_2$ ($\la 0.8$). 
This suggests that the LFs of galaxies, conditional to the group richness or group mass, are not consistent with power-law behaviour of the CLF at its  faint and bright ends.

These very low values of $T_1$ and $T_2$ may be caused by the low multiplicity of most of the
considered bins of richness, as \citet{Mamon:1987b} noticed that the values
of $T_1$ and $T_2$ are underestimated for samples of groups with fewer than
30 galaxies. We checked this effect by building mock samples of groups, where
the galaxy luminosities are drawn from a \citet{Schechter:1976} LF, with
parameters taken from Blanton et al. (2003): faint-end slope $\alpha=-1.05$
and $L/L^* > 0.24$ (given our limiting absolute magnitude of \Mabs$\,\leq
-19.57$). In each bin of richness, we built the groups with the same
distribution of multiplicity. For each sample of groups, we computed the
fraction of groups with high (\deltam$\,\ge 2$) and low (\deltam$\,\leq 0.3$)
magnitude gaps, as well as the values of $T_1$ and $T_2$. We computed the $5^{\rm th}$
and $95^{\rm th}$ percentiles of these values for $1000$ random samples built in this
manner, as shown in red arrows in
Figs.\ref{Fig_dMag_N_Mh}c,e,g,i. This exercise led to typically $30-40$ times fewer LGGs on average (Fig.~\ref{Fig_dMag_N_Mh}c), 
typically $2-3$ times more SGGs (Fig.~\ref{Fig_dMag_N_Mh}d), and much higher values of $T_1$ ($95\%$ of the mock samples 
have $T_1$ values that are $\sim 0.6$ units higher than measured in the SDSS in every bin of richness) and $T_2$. 
Very similar results are found with $\alpha = -1.3$. This confirms that the fractions of LGGs and SGGs in the SDSS are inconsistent with a 
single Schechter form for the CLF.

The alternative is that the CLF has an additional component specific to the
BGGs, as described in Sect.~\ref{Sec_CLF}. We now go one step further and test
whether these CLFs are consistent with the distribution of gaps observed in the
SDSS.
For each bin of $M_{\rm vir}$ and $N_{\rm gals}$ shown in Figure~\ref{Fig_dMag_N_Mh}, 
we sampled the CLFs (model 6 in Table~\ref{Tab_models}) $1000$ times, building a sample of 
groups with exactly the same characteristics (i.e., number of groups and distributions of $M_{\rm vir}$ and $N_{\rm gals}$) 
as that of the sample in the bin. 
The results obtained by sampling the CLFs are shown
as the shaded areas in Figure~\ref{Fig_dMag_N_Mh}. 

The 84 percentiles of the distributions of \deltam\ in bins of $N_{\rm gals}$ (Fig.~\ref{Fig_dMag_N_Mh}a) and \MvirNoUnit\ (Fig.~\ref{Fig_dMag_N_Mh}b) from the CLF sampling 
agree with the observations. However, the 16 percentiles are higher than the ones measured from our sample.
This can be also seen in the fraction of LGGs and SGGs: while we find an agreement for the fraction of LGGs (Fig.~\ref{Fig_dMag_N_Mh}c,d),  
the sampling of the CLF underestimates the number of SGGs, and the discrepancy increases with decreasing $N_{\rm gals}$ and halo mass (Fig.~\ref{Fig_dMag_N_Mh}c,d).
Moreover, the CLF sampling results in $T_1$ and $T_2$ values that are, respectively, $0.02 - 0.1$ (Fig.~\ref{Fig_dMag_N_Mh}g,h) and $0.08 - 0.2$ (Fig.~\ref{Fig_dMag_N_Mh}i,j) lower than observed. 

The discrepancies between statistics might be due to the fact that the \deltam\ values of the sample in Fig.~\ref{Fig_dMag_N_Mh} are computed within $0.5\,r_{\rm vir}$, 
while we are sampling the CLFs that were fitted to all galaxies within $2\,$\Rvir\ (see Sect.~\ref{Sec_CLF}). 
However, the statistics of gap obtained with SBGGs defined within $2\,$\Rvir\ 
are also discrepant with those predicted by these CLFs (see Sect.~\ref{Sec_Rrvir}).
In addition, if we consider the CLF of galaxies within $0.5\,$\Rvir, we get similar results and are still not able to reproduce the gap statistics.

We also compared our results with the $T_1$ and $T_2$ values from studies by \citet[][hereafter DG+12]{DiazGimenez.etal:2012}, 
\citet[][LS06]{Loh.Strauss:2006}, and \citet[][LOM10]{Lin.etal:2010}. Our $T_1$ values are much lower than those found by 
LS06 and LOM10. As indicated in Fig.~\ref{Fig_dMag_N_Mh}h, LS06 found $T_1 = 0.75 \pm 0.1$ for a sample of nearby rich SDSS clusters, while we obtain 
$T_1 \sim 0.4$ for groups with halo masses similar to those of their sample. LOM10 also found much higher values of $T_1$: $0.84 \pm 0.01$ and $0.70 \pm 0.1$ 
for low and high luminosity SDSS clusters, respectively. These comparisons should be taken with caution, because LS06 and LOM10 consider much wider bins of group mass that causes blurring between the locations of the central and satellite components of the CLFs. Interestingly, LOM10 obtain a higher $T_1$ for low luminosity clusters compared to high luminosity clusters, 
in contrast to our results, which show that $T_1$ increases with increasing halo mass.
DG+12 found $T_1 = 0.51 \pm 0.06$ for a complete sample of compact groups drawn from the 2MASS catalogue, which is lower than the values obtained by the previous authors, but still higher
than ours ($T_1 \sim 0.4$, Fig.~\ref{Fig_dMag_N_Mh}g).
80 per cent of their groups  have richness $N_{\rm gals}=4$, therefore these compact groups span a narrower range of halo masses than the cluster samples of LS06 and LOM10, yet a wider range than our own mass bins. This explains that their $T_1$ and $T_2$ values are lower than other published studies, but still larger than ours.
On the other hand, as shown in Fig.~\ref{Fig_dMag_N_Mh}i,j, the values of $T_2$ obtained by LS06 ($T_2 = 0.86 \pm 0.1$), LOM10 ($0.77, 0.79 \pm 0.01$), and DG+12 ($0.70 \pm 0.06$) are in better agreement with our results ($T_2 \sim 0.75 - 0.80$). 

\subsection{Gap statistics from a semi-analytical model}

We compared our results with predictions from the semi-analytical model (SAM) of \citet{Henriques.etal:2015}, run on the Millennium-II simulations \citep{BoylanKolchin.etal:2009}. We extracted the snapshot corresponding to $z = 0$ from the {\tt Henriques2015a..MRIIscPlanck1} table in the Virgo--Millennium database of the German Astrophysical Virtual Observatory (GAVO\footnote{\url{http://gavo.mpa-garching.mpg.de/portal/}}).

From the simulation box extracted from GAVO, we built a mock flux-limited,
SDSS-like sample of groups and galaxies, following the steps of \citet{Duarte.Mamon:2014}. Since 
the simulation box is not large enough to produce the SDSS-like group catalogue, 
we replicated the simulation box along the three Cartesian coordinates, then
placed an observer at some position and mapped the galaxies on the sky.
The absolute magnitudes in the 
$r$-band (including internal dust extinction) were converted to apparent magnitudes, and the flux limit of the
Main Galaxy Sample of the SDSS, ${m_r < 17.77}$, was applied.
We added errors of $0.2$~dex to the group halo masses and of $0.08\,\hbox{mag}$ 
to the absolute magnitudes, as determined by \citet{Duarte.Mamon:2015} for the galaxies in the SDSS/MGS.
We then selected the galaxies and groups from the mock catalogue
following the same selection criteria that we had applied to the observations 
and presented in Sect.~\ref{Sec_sample}.

We obtained the distribution of \deltam, fractions of LGGs and SGGs, as well as the 
$T_1$ and $T_2$ values as a function of halo mass (solid black lines in the right panels of Fig.~\ref{Fig_dMag_N_Mh}).
The median value of the \deltam\ distribution from the SAM is $\sim 0.3\,$mag lower (Fig.~\ref{Fig_dMag_N_Mh}b) and 
the SAM predicts fewer LGGs (Fig.~\ref{Fig_dMag_N_Mh}d) than observed.
On the other hand, the fraction of LGGs is compatible with the observations, at least for groups with \Mvir$\, \gtrsim 13.3$.

We observe much lower $T_1$ values than those predicted from the SAM (Fig.~\ref{Fig_dMag_N_Mh}h). 
Although  higher than those computed from observations, the $T_1$ values from the SAM are still 
lower than those obtained by LOM10 for groups with \Mvir$\, \lesssim 13.7$, but
they are in agreement with LOM10 and LS06 for \Mvir$\,> 13.7$.
But, again, the comparison with LOM10 and LS06 is not a fair one since they consider much wider bins of group mass than we do.
he discrepancies between the $T_2$ values from the SAM and SDSS sample are much smaller than 
that of $T_1$, with lower observed $T_2$ values than those from the SAM (Fig.~\ref{Fig_dMag_N_Mh}j).

\subsection{Magnitude gap and groups with two central galaxies}
\label{Sec_twoCentrals}

The fractions of SGGs from the CLF sampling are lower than observed, indicating that the SBGGs are brighter than predicted by 
the satellite CLF. Therefore, one may ask whether, in some SGGs, both the BGG and the SBGG come from the same distribution. 

Indeed, when groups merge, the magnitude gap will be altered at the time of the group merger and again when the two brightest (and usually most massive) galaxies merge. 
Assume that group 1 is more massive than group 2, and that its BGG is more luminous and massive than that of group 2.  Suppose also that the two groups merge before the time when the SBGG and BGG of group 1 merge together. This would suggest that, before the group merger, the BGG of group 2 is more massive, hence more luminous than the SBGG of group 1. Thus, when the groups merge, the magnitude  gap will suddenly decrease to a lower value. At the same time, one can consider that the merged group has two centrals. Once the original BGG of defunct group 2 merges into the BGG of  the merged group (that of the original group 1), after orbital decay by dynamical friction, the group will have a single central galaxy, and the gap will now correspond to the luminosity ratio of the galaxy created by the merger of the two BGGs with the SBGG of the original group 1. This gap will be larger than the original gap of group 1 before the group-group merger. This simple scenario is complicated by many features such as additional  group-group and galaxy-galaxy mergers. But, to first order, one expects that groups are more likely to have 2 centrals if they have small gaps.

We therefore propose a model where \emph{some} groups have \emph{two} galaxies whose luminosities follow the CLF of central galaxies (eq.~\ref{Eq_phi_cen}).
We assume that the probability of having such a group is a function of \deltam\ as
\begin{equation}
 p_{2{\rm c}}(\Delta \mathcal{M}_{12}) = \frac{1}{2} \left[ 1 - \erf\!\left(\frac{\Delta \mathcal{M}_{12} - \mu}{\sqrt{2}\, \sigma} \right) \right] \ .
 \label{Eq_p2c}
\end{equation}

We then repeat the exercise of sampling the CLFs allowing some groups to have two central galaxies. One may think that we would then need to recompute the CLF, but while the central and satellite CLFs will change, the total CLF will not, as suggested by the lack of sensitivity, seen in Figure~\ref{Fig_dMag_N_Mh_allModels}, of the gap statistics to the choice of free parameters in the analytical fits of the CLF.

\begin{figure}
\centering
\begin{tabular}{c}
 \Large{$\bm{R_{\rm max} = 0.5\,r_{\rm vir}}$} \\ 
 \includegraphics[width=0.99\hsize]{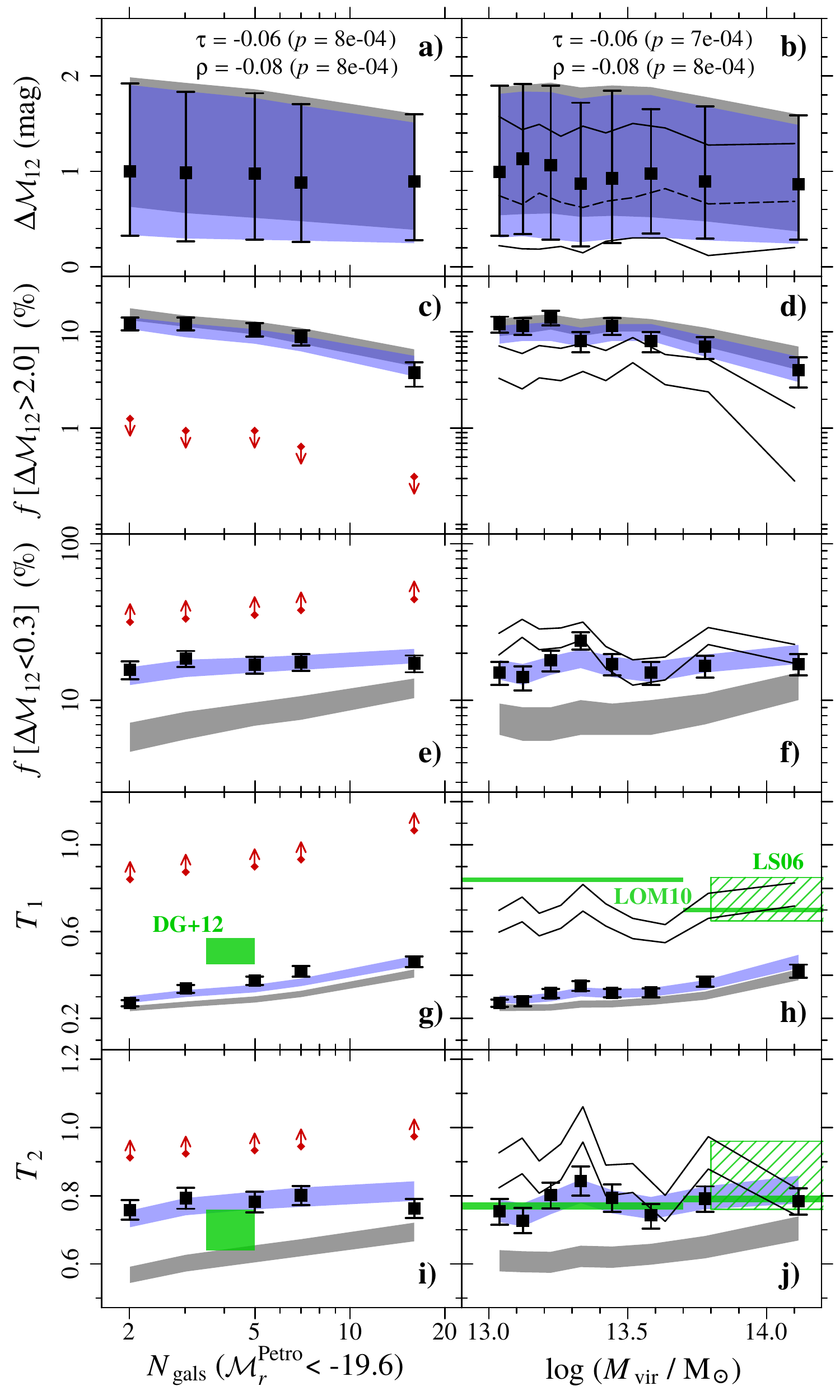} 
\end{tabular}
\caption{Same as Fig.~\ref{Fig_dMag_N_Mh}, but showing gap statistics when we allow some groups to have ``two central galaxies'' according to Eq.~(\ref{Eq_p2c}) with $\mu=0.4$ and $\sigma=0.2$ (\emph{shaded blue areas}). 
} 
\label{Fig_dMag_N_Mh_2}
\end{figure}

We find that the observed gap statistics are now very well reproduced when we sample the CLF with equation~(\ref{Eq_p2c}) for ${\mu = 0.4}$ and ${\sigma = 0.2}$ (Fig.~\ref{Fig_dMag_N_Mh_2}).
Therefore, it seems that in groups with small gaps, the SBGG luminosities follow the same CLF as the BGG, i.e., they have two central galaxies. 
This can be a consequence of mergers of groups.
Note that this functional form does not have a physical motivation; it merely provides an adequate description of the observations.

%

\subsection{Gap statistics versus \Rmax}
\label{Sec_Rrvir}

\begin{figure*}
\centering
\begin{tabular}{cc}
 \Large{$\bm{R_{\rm max} = 1.0\,r_{\rm vir}}$} & \Large{$\bm{R_{\rm max} =\,2.0\,r_{\rm vir}}$}  \\
 \includegraphics[width=0.49\hsize]{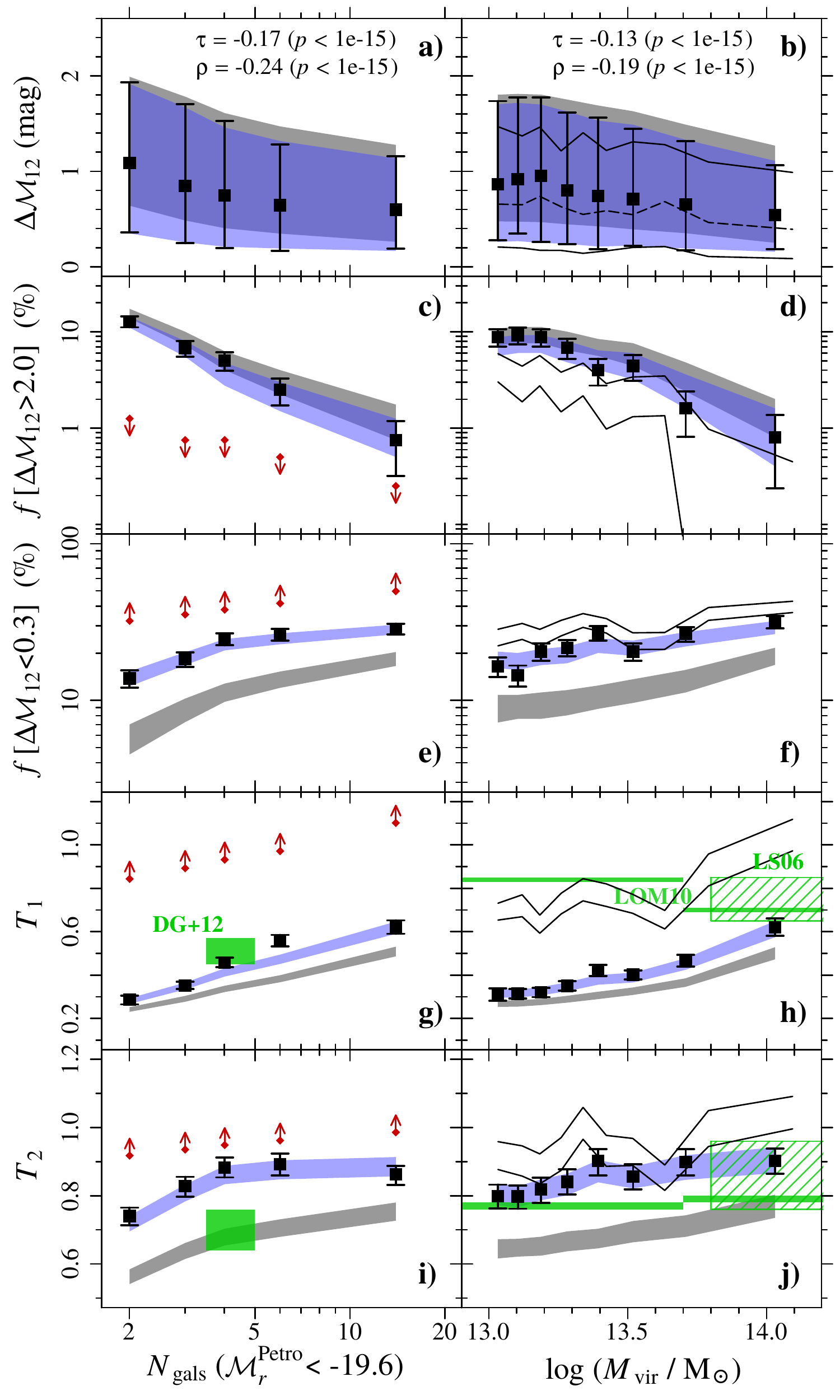} &
 \includegraphics[width=0.49\hsize]{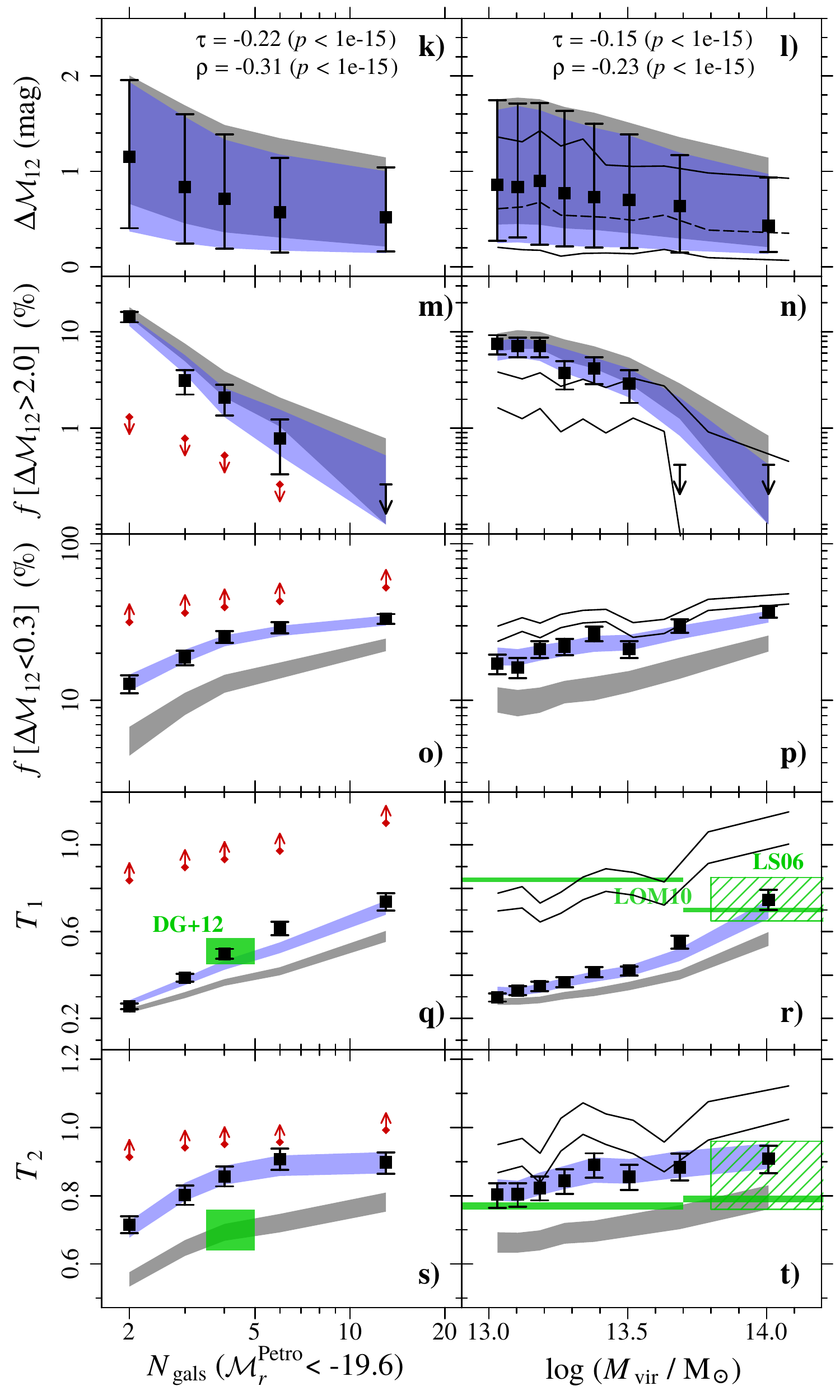}  \\
\end{tabular}

\caption{Same as Fig.~\ref{Fig_dMag_N_Mh_2},
but
changing the maximum radius, \Rmax, allowed for obtaining the SBGG to $R = 1.0\,r_{\rm vir}$ (\emph{left}) and $2.0\,r_{\rm vir}$ (\emph{right}).} 
\label{Fig_dMag_N_Mh_3}
\end{figure*}

Since the SBGGs can be further than $0.5~r_{\rm vir}$ from the BGG, the statistics of magnitude gap might
depend on the maximum radius, \Rmax, allowed to obtain the SBGG.
To investigate how different \Rmax\ affects the statistics of the magnitude gap, we now repeat the analysis for the sample defined within 
1 and $2\,$\Rvir\ (see Sect.~\ref{Sec_samples_r} and 
Table~\ref{Tab_samples}). The results are presented in Fig.~\ref{Fig_dMag_N_Mh_3}.

%

As was the case for \Rmax$\, = 0.5\,$\Rvir,  we are not able to reproduce the observed gap statistics when we assume that all groups have only one central galaxy (grey shaded regions).
However, when we allow some groups to have two central galaxies,
as described in Sect.~\ref{Sec_twoCentrals},  the observed gap statistics are reproduced by those obtained by sampling the CLF, allowing one or two centrals (blue shaded regions).

\begin{figure}
\centering
 \includegraphics[width=0.99\hsize]{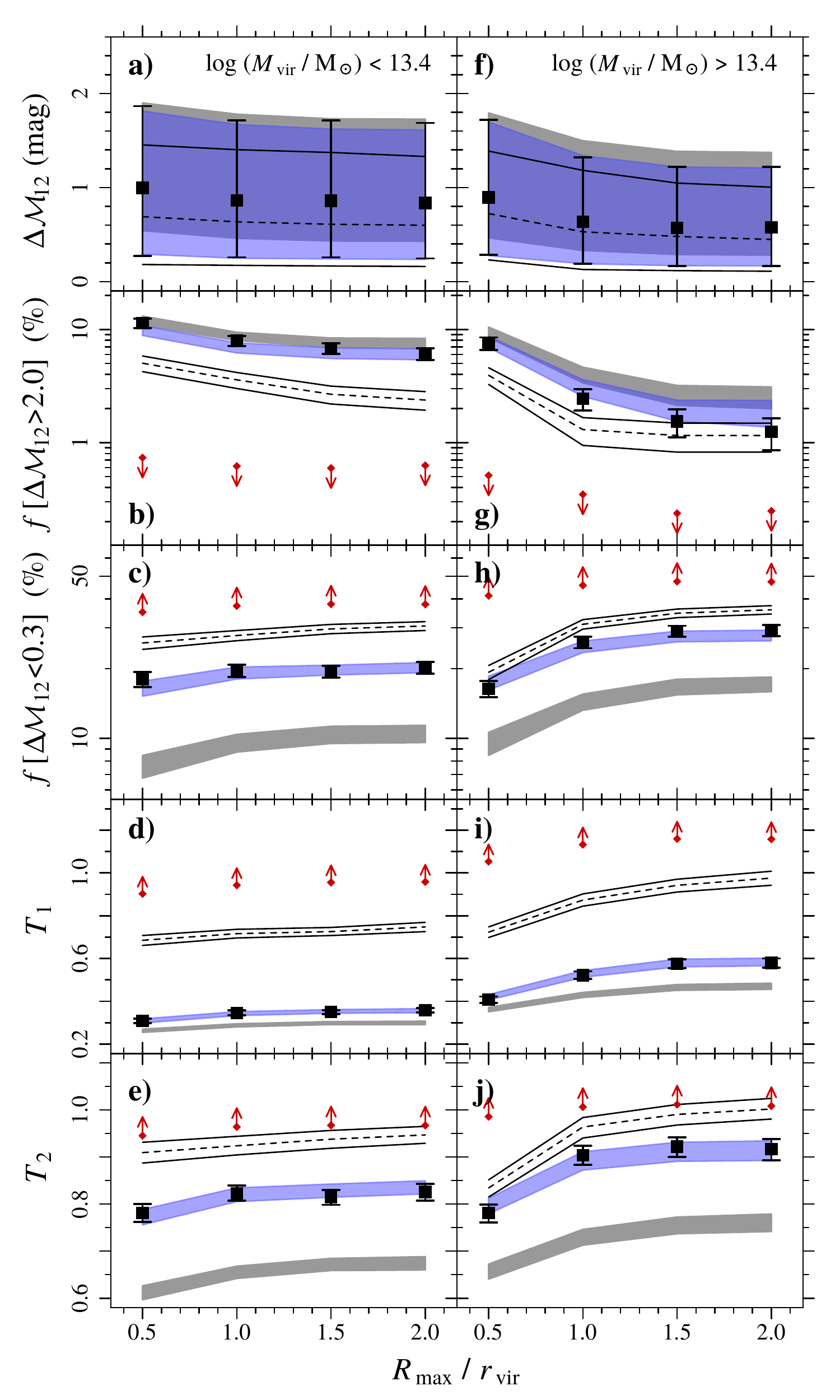} 
\caption{Statistics of the magnitude gap versus the maximum radius, \Rmax, allowed for obtaining the SBGG:
magnitude gap ({\bf a, b}); fractions of large gap groups  (\deltam~$ > 2.0$, {\bf c,d}) and small gap groups (\deltam~$ < 0.3$, {\bf e,f}); $T_1$ ({\bf g,h}); and $T_2$ ({\bf i,j}).
The \emph{left} and \emph{right panels} show the results for groups with \Mvir~$< 13.4$ and $> 13.4$, respectively.
The shaded areas represent the $16\%$ and $84\%$ when we assume that all groups have only one central (\emph{shaded gray}) and when we allow some groups to have two central galaxies according to Eq.~(\ref{Eq_p2c}) with $\mu=0.4$ and $\sigma=0.2$ (\emph{shaded blue areas}).
The \emph{dashed} and \emph{solid black lines} indicate the values and uncertainties estimated from the semi-analytical model by \citet{Henriques.etal:2015}.} 
\label{Fig_stats_Rmax}
\end{figure}

In Fig.~\ref{Fig_stats_Rmax}, we show a more complete picture of how the magnitude gap statistics vary with the maximum distance allowed to find the SBGG. 
We present the magnitude gap, the fractions of groups with \deltam~$ > 2.0$ and $< 0.3$ mag, as well as the $T_1$ and $T_2$ values as a function of \Rmax\ in two bins of halo mass 
(\Mvir~$< 13.4$ and $> 13.4$). 

For low mass groups, the gap statistics vary very slightly with \Rmax. 
For \Rmax$/r_{\rm vir}=0.5$ to $2$, the median \deltam\ value decreases $0.14\,$mag, the fraction of LGGs varies from $11.4\%$ to $8.0\%$, SGGs from $18\%$ to $20\%$, 
and $T_1$ ($T_2$) from 0.31 (0.78) to 0.35 (0.82).
On the other hand, for \Mvir$\,> 13.4$, there is a strong variation from \Rmax$/r_{\rm vir}=0.5$ to 1.0: 
we observe a decrease of $0.25\,$mag in the median \deltam\ value, and the fractions of LGGs and SGGs vary from $7.6\%$ to $2.4\%$ and $16\%$ to $26\%$, respectively.
The value of $T_1$ ($T_2$) varies from 0.41 (0.78) to 0.52 (0.90). 
For \Rmax$\, > r_{\rm vir}$, all the quantities vary very little with \Rmax.
All these results are very well reproduced when we allow groups with 2 central galaxies.

The predictions from the SAM here are strikingly different from the observations, in particular for low-mass groups.
Finally, our results remain inconsistent with the values expected for a single \citeauthor{Schechter:1976} LF for all \Rmax\ values.

%

\section{Discussion}
\label{Sec_discussion}

\subsection{The bright end of the conditional luminosity functions}
\label{Sec_CLF_discussion}

The differences that we find between our CLFs and those of Y08 can arise
from differences in the sample definition. 
Y08 include galaxies up to $z = 0.2$, while our sample contains only groups at $z < 0.07$. 
In addition, their analysis was based on SDSS-DR4, while our sample was defined from DR7, and we use DR12 photometry (which is thought to be more accurate for massive ellipticals with shallow surface brightness profiles).
Y08 used a flux-limited sample, while we use a doubly complete one, and they 
applied an evolution correction to the magnitudes, while we do not (although we 
estimate the correction from our sample, these estimates are not precise).
Finally, the methods used to obtain the best-fit CLF are different: while Y08 appear to have applied $\chi^2$ minimization to the binned data, we use MLE with no binning.

We obtain steeper $\alpha$ vs. \MvirNoUnit\ relations than Y08 (Fig.~\ref{Fig_model_sat}). 
From \Mvir$\,= 13$ to $15$, we find that $\alpha$ decreases (steepens) from $\sim -0.4$ to $-1.8$ (model 2), while $\alpha$ varies between $\sim -0.9$ to $-1.6$ for the Y08 model. 
This extra steepening of the faint-end slope in our fits may be a consequence of our cleaning the sample to ensure the correct identification of the SBGG (see Sect.~\ref{Sec_fiber_col}).
Indeed, if we do not clean the sample to ensure that SBGGs are correctly identified and redo our CLF fits (Fig.~\ref{Fig_fitCLF_ap} and Table~\ref{Tab_models_ap}) (Fig.~\ref{Fig_model_sat_ap}), we obtain a shallower relation between $\alpha$ and group halo mass.

\begin{figure}
\centering
 \includegraphics[width=0.99\hsize]{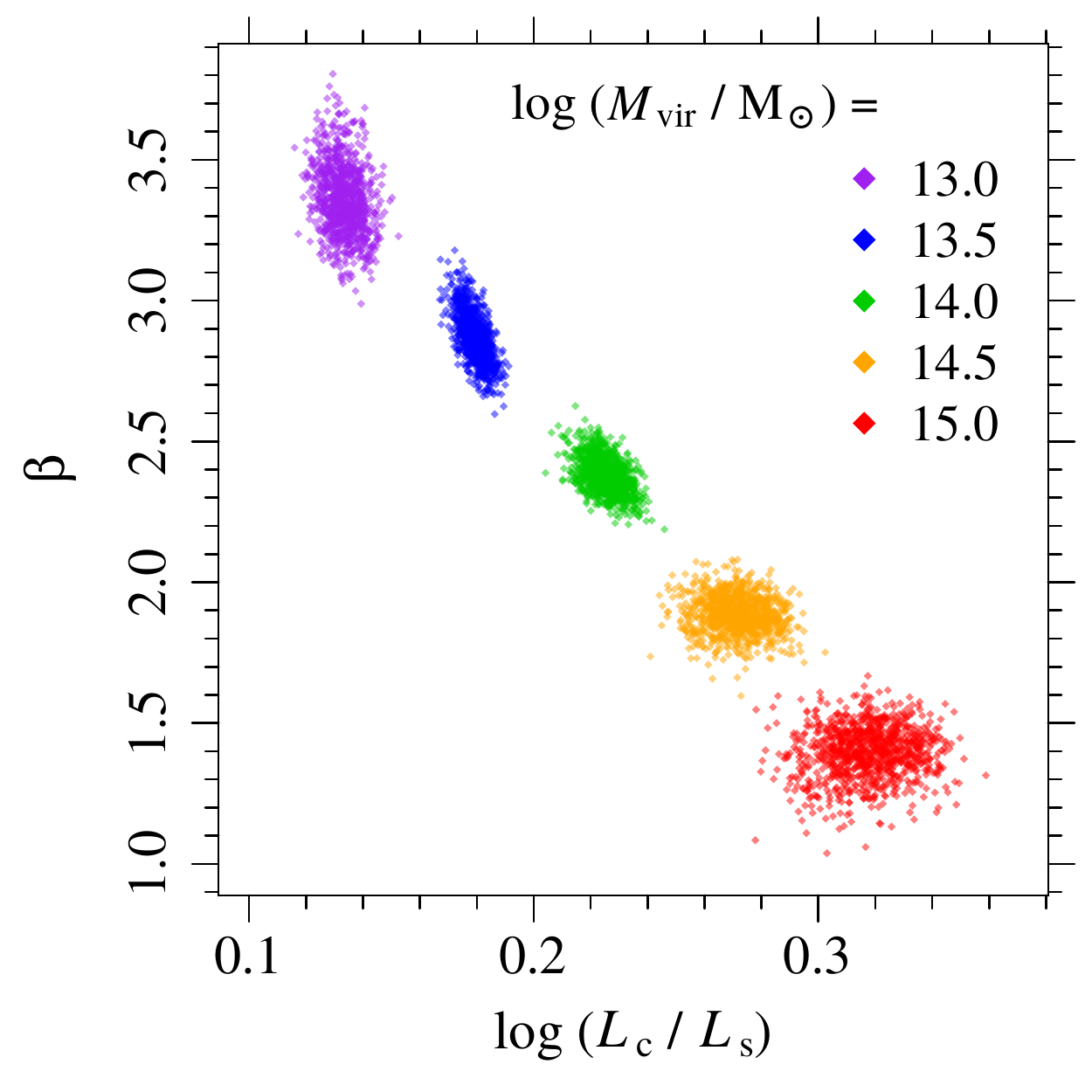} 
\caption{Illustration of the anti-correlation between the bright-end shape of the satellite conditional luminosity function and the characteristic luminosity (relative to the characteristic central luminosity) for different halo masses. The figure shows the best-fit parameters of model 6 obtained by bootstraping the group sample $1000$ times.} 
\label{Fig_beta_LsLc}
\end{figure}

The characteristic luminosity of the satellite CLF (relative to the characteristic central luminosity) increases with \MvirNoUnit, 
while the bright-end cut-off $\beta$ decreases with increasing halo mass (Fig.~\ref{Fig_model_sat}). 
Since high $\beta$ values can be compensated by low $\log (L_{\rm c} / L_{\rm s})$ ratios, one might 
ask whether there is degeneracy between these two parameters.  
In Fig.~\ref{Fig_beta_LsLc} we show that at a fixed and low halo mass, there is indeed an anti-correlation between these two parameters, but it is much smaller than the variation that we  observe 
between low and high-mass haloes (\Mvir$\,=13 - 15$).

%

\subsection{Magnitude gap statistics versus group richness and halo mass}

When comparing our statistics on the magnitude gaps with those of previous authors, one should be careful to note that two effects alter these statistics. First, samples (bins) of larger richness will naturally lead to smaller gaps. Second, the Tremaine-Richstone statistics $T_1$ and $T_2$, which measure departures from one-component  LFs or CLFs,  will be less efficient in wide ranges of group halo masses. Indeed,  since the CLF depends on group mass with more luminous characteristic central and satellite luminosities  at higher group mass, wider mass bins will wash out the separation between centrals and satellites, and will lead to higher values of $T_1$ and $T_2$.

Using semi-analytical models of galaxy formation, \citet{Dariush.etal:2007}, \citet*{DiazGimenez.etal:2008}, and \citet{Farhang.etal:2017} have shown that major progenitors of LGGs assemble half their mass earlier than do regular groups of similar final mass, as previously shown by \citet{DOnghia.etal:2005} using hydrodynamical cosmological simulations. Since the major progenitors of present-day massive haloes assembled their mass later than those of lower mass haloes \citep{vandenBosch:2002}, one can conclude that early mass assembly is linked to both low final-mass haloes and to present-day LGGs, hence to a possible anti-correlation of magnitude gap with halo mass. Physically, given the flattening of the stellar mass versus halo mass relation at high halo masses \citep[e.g.][]{Yang.etal:2009}, the dynamical friction times should be longer in high mass clusters than in low mass groups, so if the wide magnitude gaps of LGGs are caused by mergers, one  indeed expects an anti-correlation of magnitude gap with group halo mass.  Our results confirm this anti-correlation of magnitude gap with group mass (Fig.~\ref{Fig_dMag_N_Mh}b): the negative trend is significant ($p = 7-8 \times 10^{-4}$). The most massive LGG in our sample (defined within \halfRvir) has \Mvir~$= 14.2$, while the most massive systems among the SGGs have halo masses as high as \Mvir~$ = 15.1$.

Applying a semi-analytical model of galaxy formation \citep{deLucia.Blaizot:2007} to the
Millennium simulation \citep{Springel.etal:2005}, 
\citet{DiazGimenez.etal:2008} predict that only $5.5\%$
of the systems with \Mvir~$> 13.7$ are LGGs, in agreement with the fraction
that we find in the same mass range, which is $5.6 \pm 1.2\%$ ($21$ out of
$377$ groups. \citet{Tavasoli.etal:2011} also find an anti-correlation between the fraction of LGGs and group mass for both a sample of
SDSS groups and from the outputs of a semianalytical model
\citep{Bower.etal:2006} run on 
the Millenium simulation.
In addition, our CLFs also lead to a decreasing \deltam\ with
increasing halo mass (see Sect.~\ref{Sec_TRstat}, in particular
Fig.~\ref{Fig_dMag_N_Mh}b).

On the other hand, this anti-correlation of gap with halo mass is in contradiction with the results by
\citet{Hearin.etal:2013}, who found that haloes with richness between $12$
and $18$ and \deltam$\,\leq 0.2\,\hbox{mag}$ are less massive than haloes of the same
richness with \deltam$\,\geq 1.5\,\hbox{mag}$. They reached this conclusion by analyzing a
mock galaxy catalogue built by subhalo abundance matching of galaxy
luminosities to the subhalo masses of the Bolshoi cosmological $N$-body simulations
\citep*{Klypin.etal:2011}.
If we repeat their selection by limiting our
sample to groups that have $12$ to $18$ galaxies brighter than
\Mabs~$=-19.5$, we find that the halo masses of groups with \deltam$\,\leq 0.2$ are compatible with those of groups with $\geq 1.5\,\hbox{mag}$, in disagreement with their results.
This disagreement suggests that, while subhalo abundance matching reproduces the clustering
of galaxies in a large range of masses and redshifts \citep*{Conroy.etal:2006}, it
cannot match the tail end of the CLF.
Part of this discrepancy may also lie in \citeauthor{Hearin.etal:2013}'s use
of a group Friends-of-Friends group finder to extract their groups that was
non-optimal in its adopted linking lengths \citep{Duarte.Mamon:2014} and non-optimal
relative to the \cite{Yang.etal:2007} group finder
\citep{Duarte.Mamon:2015}.

\begin{figure*}
\centering
 \includegraphics[width=0.99\hsize]{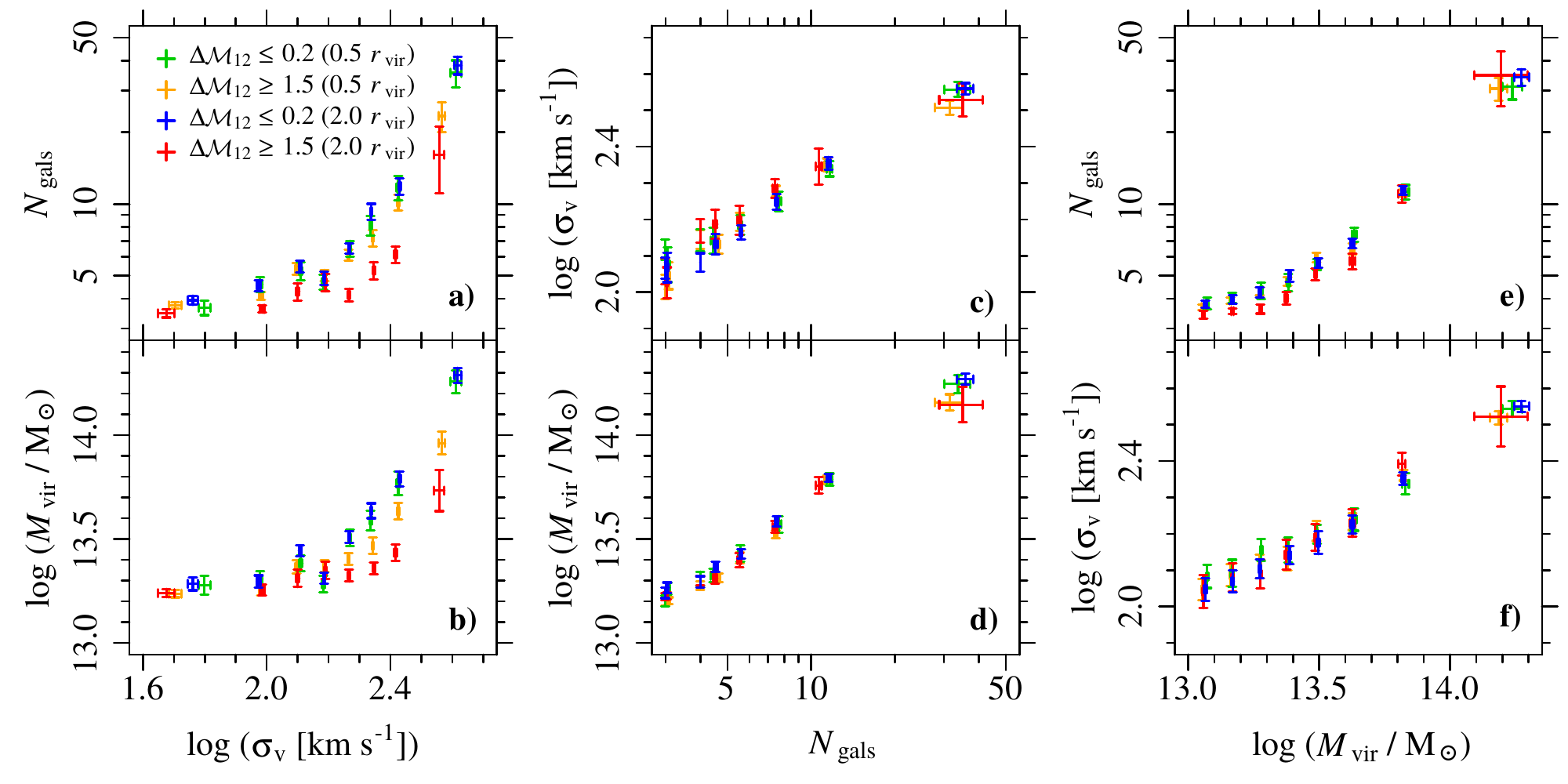} 
\caption{Relations between group velocity dispersion, mass, and richness. Small gap groups with 
\deltam$\, \leq 0.2$ and large gap groups with \deltam$\, \geq 1.5$ within \Rmax$\,=0.5\,$\Rvir\ ($2.0\,$\Rvir) are indicated as the \emph{green} (\emph{blue}) and \emph{orange} (\emph{red}) \emph{symbols}.
The error bars indicate uncertainties on the mean values from 1000 bootstraps.
} 
\label{Fig_Ngals_Mvir_sigma}
\end{figure*}

\citet{Hearin.etal:2013} also analyse SDSS groups from \citet{Berlind.etal:2006}. They find that, for fixed group velocity dispersion, groups with \deltam$\, \geq 1.5$ have fewer galaxies than groups with \deltam$\, \leq 0.2$. 
We confirm this result in Fig.~\ref{Fig_Ngals_Mvir_sigma}a for \deltam\ defined within \Rmax$\,= 2\,$\Rvir\ (red versus blue points), but not for  \Rmax$\,= 0.5\,$\Rvir\ (orange versus green points). In addition, we show in Fig.~\ref{Fig_Ngals_Mvir_sigma}b that the lower richness of LGGs is a consequence of their lower halo masses in a given $\sigma_{\rm v}$ bin. This is confirmed in Fig.~\ref{Fig_Ngals_Mvir_sigma}c,d, where we show that in bins of richness,
the masses and velocity dispersions of LGGs are compatible with those of SGGs. Therefore, the conclusion by \citeauthor{Hearin.etal:2013} that, for a fixed richness, large-gap groups are more massive than small-gap groups,  is based on their inversion of the richness versus velocity dispersion relation. In contrast, the differences between SGGs and LGGs in the direct relation of velocity dispersion versus richness are much weaker, and are even weaker in the Y08 mass versus richness relation. Moreover, in all cases any difference between SGGs and LGGs  is weakened when we limit the samples to $0.5\,r_{\rm vir}$, to such an extent that the mass versus richness relations of SGGs and LGGs are identical (orange versus green points in Fig.~\ref{Fig_Ngals_Mvir_sigma}d).

Motivated by the results of \citeauthor{Hearin.etal:2013}, \citet{More:2012} investigated analytically how the distribution of \deltam\ varies with halo mass.
He also concluded that, at fixed richness, SGGs tend to be more massive than LGGs.
However, \citeauthor{More:2012} assumes that the faint-end slope of the satellite CLFs is independent of halo mass (he adopted $\alpha = -1.17$).
As shown in Fig.~\ref{Fig_model_sat}, $\alpha$ decreases with \MvirNoUnit, and the faint-end of the CLFs of massive haloes are steeper than what was assumed by \citeauthor{More:2012}.
Although high mass haloes have more satellite galaxies, for our (and Y08's) CLFs, the probability of these satellites having low luminosities is higher. 
Therefore, the assumption of a constant $\alpha$ results in a larger fraction of bright satellites than that obtained when $\alpha$ decreases with halo mass, leading to the results obtained by \citeauthor{More:2012}.

\subsection{Gap statistics from the conditional luminosity functions: variations with \Rmax}

It is clear from our results that the single Schechter-like LF is ruled out (Figs.~\ref{Fig_dMag_N_Mh_2}), 
regardless of the $R_{\rm max}$ adopted (Figs.~\ref{Fig_dMag_N_Mh_3} and \ref{Fig_stats_Rmax}).
In addition, we are able to reproduce the observed gap statistics with CLFs corresponding to the sum of a Schechter-like and a lognormal (Sect.~\ref{Sec_CLF})
only if we allow some groups to have more than one central galaxy, with the probability of such a group being given
by eq.~(\ref{Eq_p2c}) with $\mu = 0.4$ and $\sigma = 0.2$. 
This result does not depend on the particular choice of the CLF model (see appendix~\ref{Ap_all_models}).

Moreover, our results show that the statistics of the magnitude gap depend on the adopted \Rmax\ (Fig.~\ref{Fig_stats_Rmax}). 
However,  clusters display luminosity segregation at the bright end, beyond the central galaxy, in their inner regions
\citep*{Adami.etal:1998a}, with higher giant to dwarf ratios in these cluster cores (\citealt*{Driver.etal:1998}; \citealt{Boue.etal:2008}).
It is thus surprising that the SBGGs of a large fraction ($\sim 50\%$) of our groups lie, in fact, outside \halfRvir.

\subsubsection{Are gap statistics affected by inaccurate group identifications}
\label{Sec_group_finder}

The poor identification of group members can lead to the fragmentation and merger of real-space groups. 
In the former case, the secondary fragments (which corresponds to $\sim 20\%$ of the groups obtained with the \citeauthor{Yang.etal:2007} method) have random galaxy luminosities and are preferentially located at the outskirts of the true group. But this should not affect the gap as a function of the maximum projected radius. 
On the other hand, group merging by the group finder could lead to the observed results, since a different group is included within the real group and it is more likely to be on the outskirts of the group. The BGG of the merged group would be considered the SBGG of the extracted group, and it would affect the statistics of the bright-end of the LF. 

To investigate this issue, we compared the observed statistics of the gap with those predicted by a SAM viewed in projection. 
Since these groups are perfectly extracted, the effects of poor identification should not be present and the fractions of LGGs, as well as the statistics of \deltam, 
should be the same for all samples (i.e., regardless the \Rmax\ used to define the gap). 
Yet, the comparison between Figs.~\ref{Fig_dMag_N_Mh_2} and \ref{Fig_dMag_N_Mh_3}, which presents the results for the gap defined within $0.5$, $1.0$, and $2.0\,$\Rvir,
shows that the statistics of the gap in the SAMs also depends on \Rmax. This result is more evident in Fig.~\ref{Fig_stats_Rmax}, 
where both the observations and the SAM show similar trends with \Rmax\ of LGG and SGG fractions, $T_1$, and $T_2$. 
However, the SAM predicts fewer LGGs and more SGGs than observed. In addition, the $T_1$ and $T_2$ values from the SAM are higher than those obtained from the SDSS groups. 

\subsubsection{Does the group luminosity function vary with $R$?} 
\label{Sec_LF_Rmax}

Other possibility to explain the variation of gap statistics with \Rmax\ is that the CLFs vary with distance to the group centre. 
To investigate that, we compared the distribution of luminosities of satellite galaxies at $R\leq 0.5\,r_{\rm vir}$, $0.5 < R/r_{\rm vir} \leq 1.0$, 
$1.0 < R/r_{\rm vir} \leq 1.5$, and $1.5 < R/r_{\rm vir}  \leq 2.0$. We applied the KS and AD tests, as summarized in Table~\ref{Tab_LF_R}.

We cannot rule out the possibility that the observed luminosities of satellites at $R \leq 0.5\,r_{\rm vir} $ and at
$0.5 < R/r_{\rm vir}  \leq 1.0$ come from the same distribution. On the other hand, the tests indicate that 
the distributions of satellite luminosities for samples at $0.5 < R/r_{\rm vir}  \leq 1.0$ and $1.0 < R/r_{\rm vir}  \leq 1.5$ are very different
from each other, with the outer regions lacking $L^*$ galaxies compared to the inner parts 
($p = 2 \times 10^{-12} - 3 \times 10^{-8}$).
In addition, there is a marginal evidence that 
the distributions also change from $1.0 < R \leq 1.5$ to $1.5 < R \leq 2.0$ ($p = 0.02 - 0.05$). 
These results reproduce qualitatively the variation of LFs with position, with higher giants to dwarfs ratios in the inner regions found by previous authors \citep{Driver.etal:1998,Boue.etal:2008}.

However, our analysis shows that the gap statistics for different \Rmax\ can be reproduced by a single, global CLF., allowing for 2 centrals in some groups.
Therefore, although variations of the LFs with $R/r_{\rm vir}$ seem to exist, the variations of gap 
distributions with \Rmax\ are, in fact, a consequence of number statistics.

\begin{table}
\centering
\caption{Comparison between the luminosity distributions of satellite galaxies at different distances from 
the group center. {\bf (1,3)} Radial distance limits and {\bf (2,4)} number of galaxies in the two
samples being compared; {\bf (5)} Kolmogorov-Smirnov and {\bf (6)} Anderson-Darling $p$-values.} 
\tabcolsep 6pt
 \begin{tabular}{cccccccc}
 \hline
   \multicolumn{2}{c}{Sample 1} & &  \multicolumn{2}{c}{Sample 2} & &         \multicolumn{2}{c}{$p$-values} \\
                    \cline{1-2}                    \cline{4-5}                            \cline{7-8} 
   $R$ (\Rvir) & $N_{\rm gals}$ & &  $R$ (\Rvir) & $N_{\rm gals}$ & &                KS &                 AD \\
           (1) &            (2) & &          (3) &            (4) & &               (5) &                (6) \\
  \hline                                                            
  $(0.0, 0.5]$ &           4460 & & $(0.5, 1.0]$ &           3940 & &              0.12 &               0.33 \\
  $(0.5, 1.0]$ &           3940 & & $(1.0, 1.5]$ &           2151 & & $3\times 10^{-8}$ & $2\times 10^{-12}$ \\
  $(1.0, 1.5]$ &           2151 & & $(1.5, 2.0]$ &        \ \,544 & &              0.02 &               0.05 \\
\hline
 \label{Tab_LF_R} 
 \end{tabular}
\end{table}

\subsection{Statistics of the magnitude gap and groups with two central galaxies}

Our results show that the observed statistics of the magnitude gap can be only reproduced with two-component CLFs if we allow small gap groups to have two central galaxies (Fig.~\ref{Fig_stats_Rmax}).
This scenario is expected when two groups merge, as appears to be the case in the nearby rich Coma cluster [Abell 1656] \citep{Biviano.etal:1996}, which has 
\deltam$\, = 0.21$. 

Although mergers of groups with similar halo masses (therefore with BGGs with similar luminosities) are less frequent than ``minor'' group mergers \citep*{Fakhouri.etal:2010}, the latter are less likely to fill the magnitude gap of the more massive group, i.e., the BGG of the smaller group is more likely to become a high-ranked galaxy. 

We tested if the gap statistics can be reproduced if both the BGG and SBGG luminosities are sampled from the satellite CLFs, i.e., we allow the existence of groups with no centrals. 
We find that, with the same eq.~(\ref{Eq_p2c}) with ${\mu = 0.5}$ and ${\sigma = 0.1}$, we are able to reproduce quite well the fractions of LGGs, SGGs, and $T_2$ values. However, 
the the distribution of $\mathcal{M}_1$ is wider than the observations, leading to $T_1$ values that are much higher than observed.


\section{Summary and conclusions}
\label{Sec_summary}

In this paper, we use samples of SDSS groups, extracted on the Web site of H. Yang (following the algorithm of \citealt{Yang.etal:2007}), to study 
the bright end of the galaxy CLFs. Our SDSS samples are cleaned of selection effects, and we make use of more accurate SDSS-DR12 photometry.

We found that the CLFs provided by Y08 fit well the distribution of luminosities if we allow some changes to their parameters: our best Bayesian evidence suggests that, in low-mass groups, the shape of the bright-end of the satellite component is steeper, while the ratio of characteristic satellite to central luminosities is higher compared to high-mass groups.

The statistics of magnitude gaps provide a fine test of the accuracy of the bright end of the CLFs. We first notice that these statistics depend on the maximum radius where we select the second brightest group galaxies. We also find that sampling the CLF produces too few small gap groups, regardless of our different analytical fits.  This suggests that some groups have more than one central galaxy, and indeed we find that preferentially allowing small-gap groups to have 2 centrals recovers very well all the statistics of the magnitude gap. 

Finally, we test the hypothesis of \citet{Hearin.etal:2013} that the richness-mass relation is a function of magnitude gap. We conclude that  this relation disappears when we fit mass or velocity dispersion directly to richness and when we limit our choice of the second brightest galaxies to maximum projected radii of 1 or 0.5$\,r_{\rm vir}$.






 
 
  

\section*{Acknowledgments}

MT acknowledges the financial support from CNPq (process \#204870/2014--3) and thanks the hospitality from the Institut d'Astrophysique de Paris. This research has been supported in part by the Balzan foundation via IAP. 
We acknowledge the use of SDSS data
(\url{http://www.sdss.org/collaboration/credits.html}) and {\tt TOPCAT} Table/VOTable Processing Software \citep[][\url{http://www.star.bris.ac.uk/~mbt/topcat/}]{topcat}.

\appendix

\section{Correct identification of the SBGG}
\label{Ap_cleanSBGG}

To select the group sample for our analysis of magnitude gap statistics, we ensure that the SBGGs are correctly identified, as described in \ref{Sec_fiber_col}. 
To investigate whether these selection criteria bias our sample and, therefore, the determination of the CLFs, we fitted the 
CLFs using the sample without requiring the correct identification of the SBGGs.
By applying the criteria \ref{sample_z} to \ref{sample_gap} listed in Sect.~\ref{Sec_sample} (assuming \Rmax$\,=2.0\,$\Rvir), 
and after cleaning the sample for the spectroscopic incompleteness of the BGGs and for groups near  bright stars or the edges of the survey (see Sections~\ref{Sec_fiber_col} and \ref{Sec_masks}), 
we obtain a sample of 2350 groups with a total of $13\,997$ satellites within $2.0\,$\Rvir.

\begin{figure}
\centering
\includegraphics[width=\hsize]{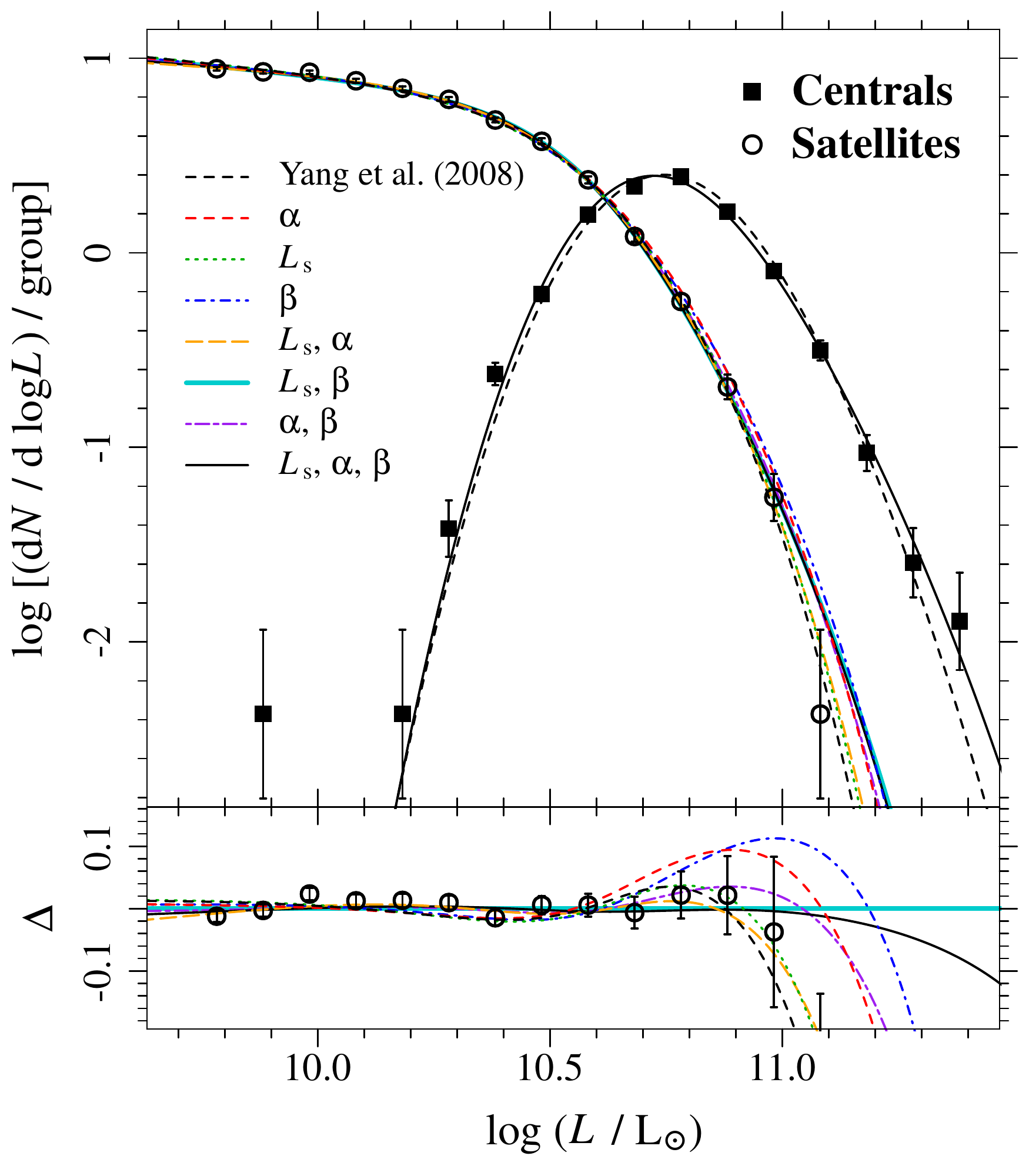}
\caption{Same as Fig.~\ref{Fig_fitCLF_sbggClean} (with the same notation), but for the sample for which we do not require the correct identification of SBGGs.} 
\label{Fig_fitCLF_ap}
\end{figure}

\begin{figure}
\centering
\includegraphics[width=\hsize]{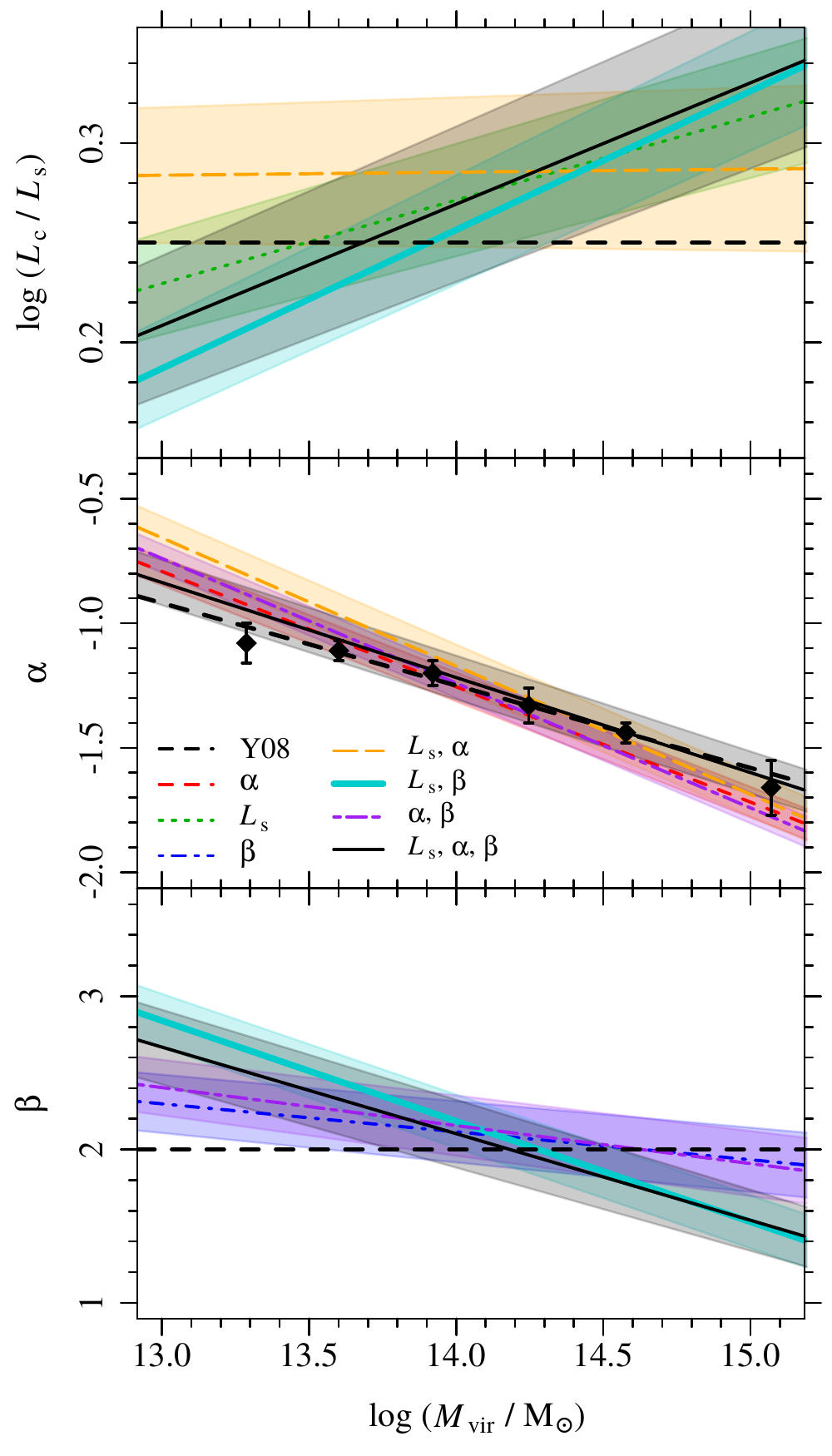}
\caption{Same as Fig.~\ref{Fig_model_sat} (with the same notation), but for the CLFs shown in Fig.~\ref{Fig_fitCLF_ap}.} 
\label{Fig_model_sat_ap}
\end{figure}

In Figure~\ref{Fig_fitCLF_ap}, we show the LF of central and satellites galaxies, computed as the best-fit CLFs averaged over the halo masses of the groups in our sample. 
As in Sect.~\ref{Sec_CLF_sat}, we compared satellite CLF models with different degrees of freedom. 
The best-fit linear relations with \MvirNoUnit\ of $\log (L_{\rm c}/L_{\rm s})$, $\alpha$, and $\beta$ are shown in Fig.~\ref{Fig_model_sat_ap}.
Comparing these results with Fig.~\ref{Fig_model_sat}, we see that we obtain shallower relations of these parameters with halo mass. 

In Table~\ref{Tab_models_ap}, we show $p-$values of a KS and AD tests, as well as $\Delta$BIC and $\Delta$AIC relative to the model with the lowest BIC and AIC values. 
Differently from the results obtained with the clean sample (Table~\ref{Tab_models}), no model can be strongly rejected based on the KS and AD tests, with
the model 1 (Y08) and 3 (free $\log L_{\rm s}$) being only marginally inconsistent with the data.
On the other hand, both BIC and AIC indicate that model 6 (free $\log L_{\rm s}$ and $\beta$) is strongly preferred over the other models. 

%

%

%
\begin{table}
\centering
\caption{Comparison between models. {\bf (1)} Free parameters of the model; 
{\bf (2)} $p$-values of a Kolmogorov-Smirnov and {\bf (3)} Anderson-Darling tests;
{\bf (4)} $\Delta$BIC and {\bf (5)} $\Delta$AIC relative to the model with the lowest BIC 
and AIC (model 6) values.  Since model 1 has no free parameters, the values of BIC and AIC
correspond to $-2 \ln \mathcal{L}_{\rm s}$.}
 \begin{tabular}{lccrr}
 \hline                                               
 \multicolumn{1}{c}{Model}                               & \multicolumn{2}{c}{$p$-values} & $\Delta$BIC & $\Delta$AIC \\
                                                         &   KS-test      &       AD-test &             &             \\
 \multicolumn{1}{c}{(1)}          & \multicolumn{1}{c}{(2)} & \multicolumn{1}{c}{(3)} & \multicolumn{1}{c}{(4)}  & \multicolumn{1}{c}{(5)} \\
 \hline
 1. Yang et al. (2008)                                   &           0.06 &          0.05 &  {\it 11.6} & {\it 41.8} \\
 2. $\alpha$                                     &           0.23 &          0.18 &       20.9  &      36.0  \\
 3. $L_{\rm s}$                                          &           0.04 &          0.02 &       26.6  &      41.7  \\
 4. $\beta$                                      &           0.09 &          0.06 &       29.9  &      45.0  \\
 5. $L_{\rm s}$, $\alpha$                        &           0.58 &          0.47 &       15.8  &      15.8  \\
 6. $L_{\rm s}$, $\beta$                         &           0.21 &          0.12 & $\bm{0.0}$  & $\bm{0.0}$ \\
 7. $\alpha$, $\beta$                    &           0.39 &          0.29 &        9.6  &       9.6  \\
 8. $L_{\rm s}$, $\alpha$, $\beta$       &           0.35 &          0.29 &       17.9  &       2.8  \\
 \hline                                                      
 \end{tabular}
\label{Tab_models_ap}
\end{table}

\section{Gap statistics from different CLF models} 
\label{Ap_all_models}

\begin{figure*}
\centering
\begin{tabular}{cc}
 \Large{\hspace{1.2cm}$\bm{R_{\rm max} = 0.5\,r_{\rm vir}}$} & 
 \Large{\hspace{1.2cm}$\bm{R_{\rm max} =\,1.0\,r_{\rm vir}}$}  \\
 \includegraphics[width=0.49\hsize]{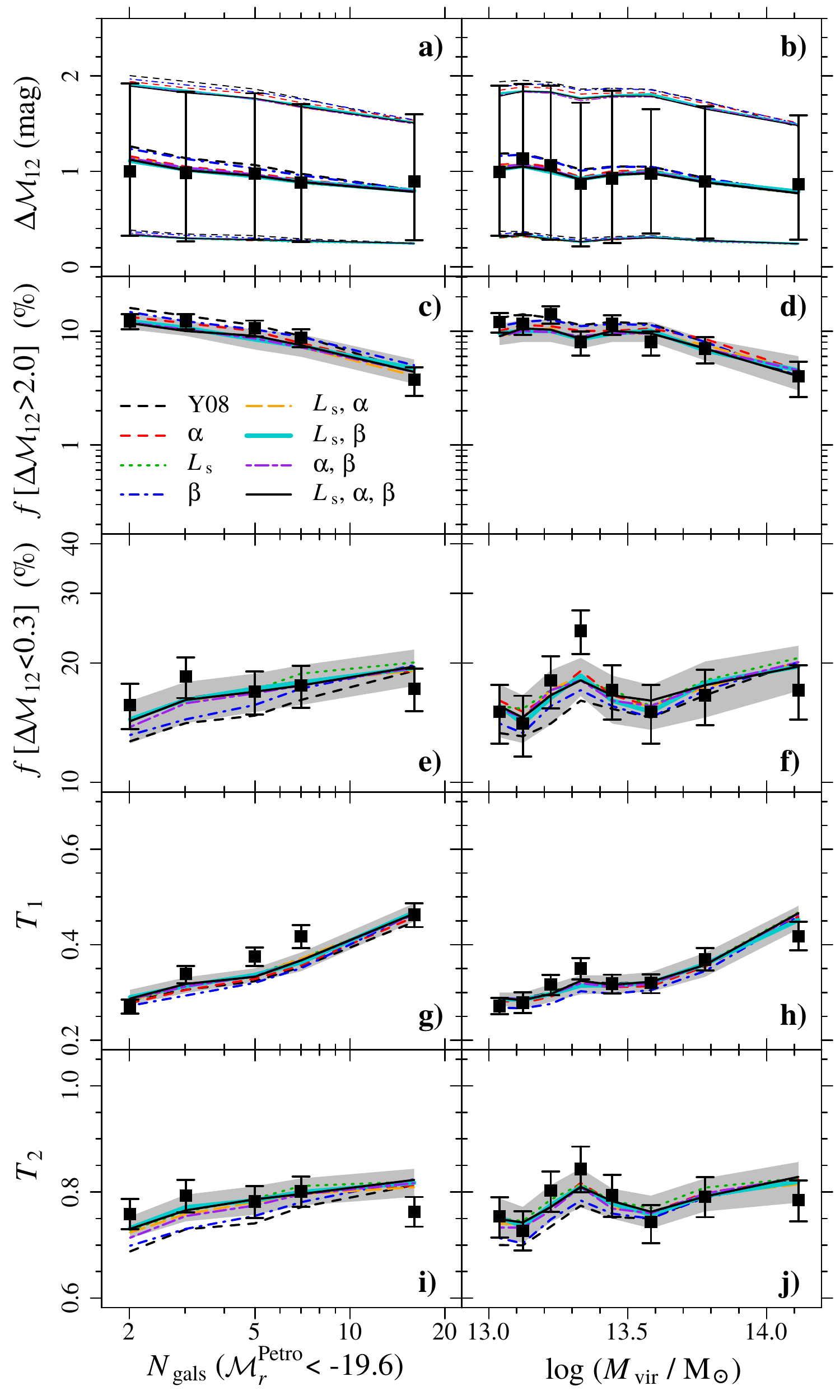} &
 \includegraphics[width=0.49\hsize]{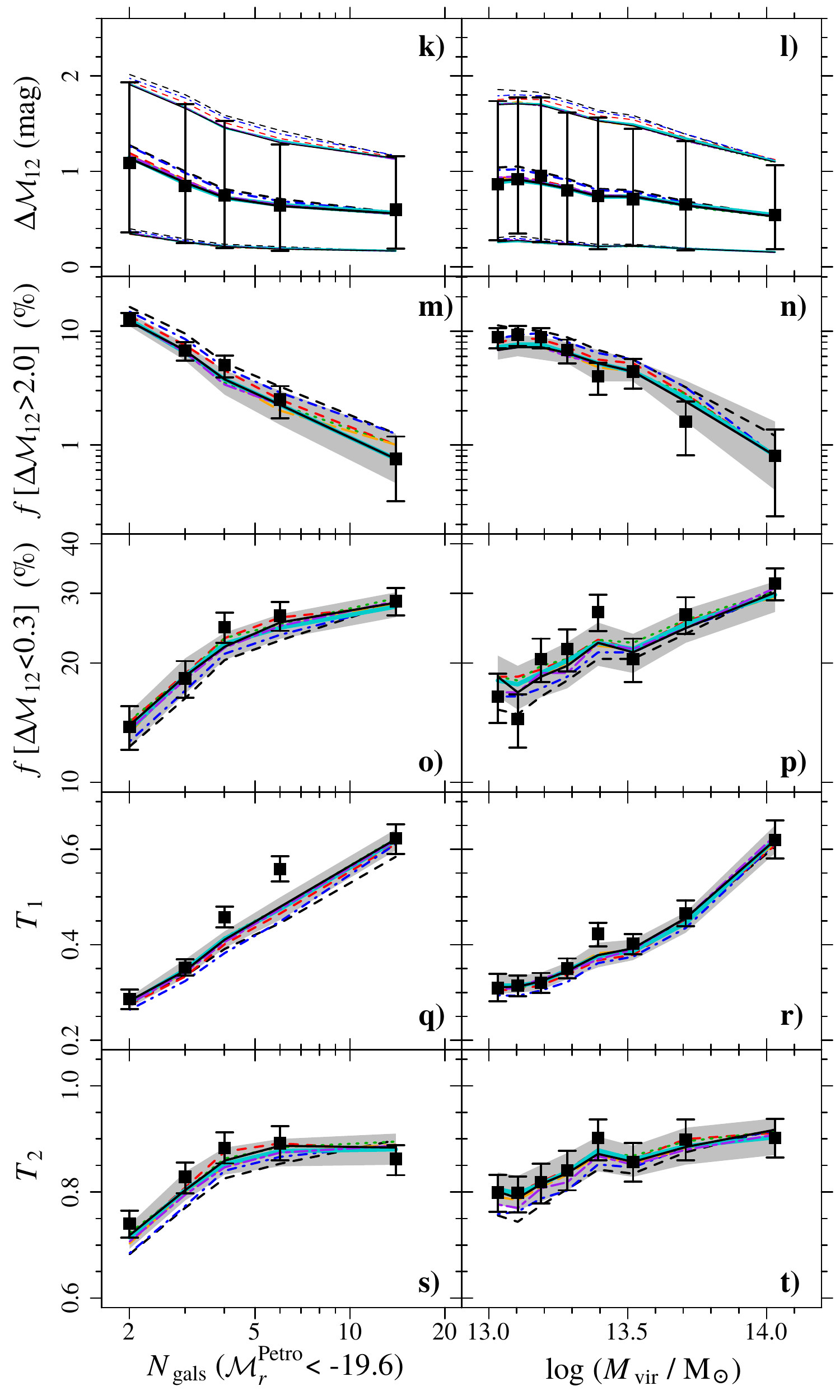}  \\
\end{tabular}
\caption{Statistics of the magnitude gap for as a function of group richness and halo mass for \Rmax$/r_{\rm vir}=0.5$ (\emph{left}) and 
$1.0\,$ (\emph{right}). Predictions from sampling different CLF models are shown: 
the best-fit models with 2 free parameters, i.e., when we fit the linear relation with \MvirNoUnit\ of 
only one of the parameters: $\alpha$ (\emph{red dashed}), $\log L_{\rm s}$ (\emph{green dotted}), 
or $\beta$ (\emph{blue dot-dashed lines}); the models with 4 degrees of freedom
 (\emph{orange long-dashed}, \emph{cyan solid}, and \emph{purple short-long dashed lines});
and the model with 6 free parameters (\emph{black solid lines}). 
The Y08 model is indicated as the \emph{black dashed lines}.} 
\label{Fig_dMag_N_Mh_allModels}
\end{figure*}

In Figure~\ref{Fig_dMag_N_Mh_allModels}, we show the gap statistics from different CLF models described in Table~\ref{Tab_models_pars}. 
Each CLF model was sampled assuming that some groups can have two central galaxies following eq.~(\ref{Eq_p2c}) with $\mu = 0.4$ and $\sigma = 0.2$, as described in Sect.~\ref{Sec_twoCentrals}. 

All models reproduce very well the observed statistics of the gap, except perhaps for models 1 (Y08) and 4 (free $\beta$). These two models lead to slightly higher fractions of LGGs (Fig.~\ref{Fig_dMag_N_Mh_allModels}c,m), lower fractions of SGGs (Fig.~\ref{Fig_dMag_N_Mh_allModels}e,f,o,p), 
lower $T_1$ (Fig.~\ref{Fig_dMag_N_Mh_allModels}q) 
and $T_2$ (Fig.~\ref{Fig_dMag_N_Mh_allModels}i,j,s,t) 
values than the other models. However, as shown in Table~\ref{Tab_models}, these 
models do not represent a good description of the data according to KS and AD tests and
BIC and AIC values. 
Although the model~3 can be also rejected based on the KS and AD tests and
BIC and AIC values, Fig.~\ref{Fig_dMag_N_Mh_allModels} shows that it provides gap statistics that are compatible with the observations and the other good models.

%

\bibliographystyle{mnras}
\bibliography{bibliografia}

\label{lastpage}

\end{document}